\documentclass[a4paper,12pt]{article}
\usepackage{amsmath,amssymb}
\usepackage[dvips]{graphicx}
\usepackage[abs]{overpic}

\numberwithin{equation}{section}
\numberwithin{figure}{section}
\numberwithin{table}{section}

\begin{document}

\begin{titlepage}
	\begin{flushright}
		UT-Komaba/11-10
	\end{flushright}
	\begin{center}
	
		\vspace*{15mm}
		
		{\LARGE\bf Signatures of low-scale string models \\at the LHC}
		
		\vspace*{20mm}
		
		{\large Manami Hashi$\,^a$ and Noriaki Kitazawa$\,^b$}
		
		\vspace{6mm}
		
		{$^a\,$\it Institute of Physics, University of Tokyo, \\
		 Komaba 3-8-1, Meguro-ku, Tokyo 153-8902, Japan \\
		 e-mail: hashi@hep1.c.u-tokyo.ac.jp} \\
		
		\vspace{3mm}

		{$^b\,$\it Department of Physics, Tokyo Metropolitan University, \\
		 Hachioji, Tokyo 192-0397, Japan \\
		 e-mail: kitazawa@phys.se.tmu.ac.jp}

	\vspace*{15mm}

\begin{abstract}

Low-scale string models,
 in which the string scale $M_{\rm s}$ is of the order of TeV
 with large extra dimensions,
 can solve the problems of scale hierarchy and non-renormalizable quantum gravity
 in the standard model.
String excited states of the standard model particles
 are possibly observed as resonances in the dijet invariant mass distribution at the LHC.
There are two properties to distinguish
 whether the resonances are due to low-scale string or some other ``new physics".
One is a characteristic angular distribution in dijet events at the resonance
 due to spin degeneracy of string excited states,
 and the other is an appearance of the second resonance
 at a characteristic mass of second string excited states.
We investigate a possibility
 to observe these evidences of low-scale string models
 by Monte Carlo simulations
 with a reference value of $M_{\rm s} = 4\,{\rm TeV}$
 at $\sqrt{s} = 14\,{\rm TeV}$.
It is shown that spin degeneracy at the dijet resonance
 can be observed by looking the $\chi$-distribution
 with integrated luminosity of $20\,{\rm fb}^{-1}$.
It is shown that the second resonance can be observed
 at rather close to the first resonance in the dijet invariant mass distribution
 with integrated luminosity of $50\,{\rm fb}^{-1}$.
These are inevitable signatures of low-scale string models.

\end{abstract}

	\end{center}
\end{titlepage}

\newpage
\setcounter{tocdepth}{1}
\tableofcontents

\section{Introduction}
	\label{sec:introduction}
	
	\hspace{5mm}
One of theoretical problems of the Standard Model (SM) is that
 it does not describe gravitational interaction in a renormalizable form.
String Theory is a strong candidate for a theory of quantum gravity.
There is another theoretical problem so-called hierarchy problem.
If the fundamental energy scale of gravitational interaction, {\it i.e.}, the Planck scale,
 is considered as a fundamental scale of the SM,
 a large hierarchy between the Planck scale
 $M_{\rm Pl} \sim 10^{19}\,{\rm GeV}$
 and the weak scale
 $M_{\rm W} \sim 10^2\,{\rm GeV}$
 can not be understood.
It is proposed that the string scale
 $M_{\rm s} = \sqrt{1/\alpha^\prime}$
 which is the fundamental scale in String Theory
 could be much lower than the Planck scale
 due to the existence of {\it large extra dimensions}
 \cite{Antoniadis:1990ew,Antoniadis:1998ig}.
This is a solution to the hierarchy problem in the framework of String Theory
 with the string scale
 $M_{\rm s} \sim {\cal O}(1)\,{\rm TeV}$.
Such string models with the low string scale are called {\it low-scale string models}
 (for review, see Ref.\cite{Lust:2009kp}).

In low-scale string models, the Planck scale is described as
\begin{equation}
	\label{eq:Plank_scale_vs_Ms}
 M_{\rm Pl}^2
  = \frac{8}{g_{\rm s}^2}
      M_{\rm s}^8 
      \frac{V_6}{(2\pi)^6} \,,
\end{equation}
where $g_{\rm s}$ is string coupling
 and $V_6$ is the volume of six-dimensional compactified space.
Note that closed strings which mediate gravitational interaction
 propagate in whole ten-dimensional space-time.
If it is assumed that $g_{\rm s}$ is small for perturbative theory,
 $V_6 M_{\rm s}^6 \sim 10^{32}$
 is required with
 $M_{\rm s} \sim {\cal O}(1)\,{\rm TeV}$.
Imagine a D$p$-brane
 whose $p$-dimensional space contains our three-dimensional space.
Open strings on the D$p$-brane give gauge symmetry,
 and the gauge coupling constant in our four-dimensional space-time is given by
\begin{equation}
	\label{eq:gauge_coupling_vs_Ms}
 \biggl(\frac{g^2}{4\pi}\biggr)^{-1}
  = \frac{2}{g_{\rm s}}
     M_{\rm s}^{p-3}
     \frac{V_{p-3}}{(2\pi)^{p-3}} \,,
\end{equation}
where $V_{p-3}$ is the volume of $(p-3)$-dimensional compactified space
 parallel to the D$p$-brane.
For appropriately large gauge coupling,
 $V_{p-3}$ should not be very large:
 $V_{p-3} M_s^{p-3} \sim 1$.
With the above condition for large $M_{\rm Pl}$,
 the volume of $(9-p)$-dimensional compactified space, $V_{9-p}$,
 should be large as much as
 $V_{9-p}M_s^{9-p} \sim 10^{32}$,
 because
 $V_6 = V_{p-3}\times V_{9-p}$.
Gravity escapes to large transverse directions to the D$p$-brane
 and the strength of gravity on the D$p$-brane becomes weak.
It has been shown in Ref.\cite{Cicoli:2011yy}
 that this type of anisotropic compactification is possible in String Theory.

A gauge symmetry U$(N)$ is realized on a stack of $N$ D-branes.
We consider models in which such stacks of D-branes relevant to the SM
 are localized in compactified space, {\it i.e.}, ``local models"
 in Ref.\cite{Lust:2008qc}.
We need, for example, four stacks of D-branes to realize
 U$(3)_{\rm color} \times $U$(2)_{\rm left} \times $U$(1) \times $U$(1)^\prime$
 gauge symmetry.
Since
 U$(N) = $ SU$(N) \times $U$(1)$,
 there is an U$(1)$ symmetry on each stack of D-branes.
The massless modes of open strings on U$(3)_{\rm color}$ branes
 are identified as gluons and an additional U$(1)_{\rm color}$ gauge boson,
 and the massless modes of open strings on U$(2)_{\rm left}$ branes
 are identified as weak bosons and an additional U$(1)_{\rm left}$ gauge boson.
The U$(1)_{\rm Y}$ gauge symmetry in the SM
 is an independent linear combination of four U$(1)$ symmetries.\footnote{Remaining three U$(1)$ symmetries are usually ``anomalous",
  and these gauge bosons are massive.}

The chiral matter for the SM is realized by open strings
 whose two ends attach on two different stacks of D-branes
 intersecting with each other \cite{Berkooz:1996km}.
Space of the intersection between the two different stacks
 contains our three-dimensional space
 and the open strings are localized on that space.
For example, left-handed quark doublets $Q_\mathrm{L}$ are realized
 by the massless modes of open strings
 between U$(3)_\mathrm{color}$ and U$(2)_\mathrm{left}$ branes,
 right-handed up-type quarks $\bar{U}_\mathrm{R}$ are realized
 by the massless modes of open strings
 between U$(3)_\mathrm{color}$ and U$(1)$ branes, and so on.

In addition to the SM particles, there are many massive states:
 ``string excited states" as massive modes of open strings,
 Kaluza-Klein (KK) states of the SM particles and string excited states,
 and various closed string states.

In this paper, we explore a possibility
 to observe signatures of low-scale string models
 in dijet events at the LHC.
As it is reviewed in the rest of this section,
 there are some model-independent features in parton two-body scattering amplitudes.
We investigate dijet events including contributions of string models
 by Monte Carlo simulations.
We concentrate on the following two distinct properties.
\begin{itemize}
 \item A resonance in the dijet invariant mass distribution by ``string excited states"
  consists many degenerated states with various spins.
 An analysis of the dijet angular distribution at the resonance must be very important
  to have a signature of low-scale string models
  \cite{Anchordoqui:2009mm,Anchordoqui:2008di,Kitazawa:2010gh}.
 \item String $n$th excited states have masses of
  $M_n = \sqrt{n} M_\mathrm{s}$.
  A second resonance
  at a place $\sqrt{2}$ times far from the place of the first resonance
  in the dijet invariant mass distribution,
  is a signature of low-scale string models.
\end{itemize}

In the rest of this section,
 we discuss the model independency
 of predictions to parton two-body scatterings at the LHC
 by low-scale string models \cite{Lust:2008qc}.
We begin with summarizing generally spectra of low-scale string models
 in four-dimensional space-time.

The spectra of excited states of open strings
 can be understood from open-string two-body scattering amplitudes
 which are calculated by the world-sheet superconformal field theory in flat space-time.
If we consider a D$3$-brane
 where the world volume coincides with our four dimensional space-time,
 the open-string amplitudes depend only on the string scale $M_\mathrm{s}$,
 group theoretical factor and gauge coupling constant,
 and do not depend on the details of model buildings
 such as the way of compactification and configuration of D-branes.
String effects in the open-string two-body scattering amplitudes
 are described by ``string form factors"
 which are functions of the Mandelstam variables $s$, $t$ and $u$ (with $s+t+u = 0$)
 \cite{Cullen:2000ef}.
A typical form of the form factor function is
\begin{equation}
	\label{eq:V-function}
 V(s,t,u)
  = \frac{\Gamma(1-s/M_\mathrm{s}^2)\Gamma(1-u/M_\mathrm{s}^2)}
              {\Gamma(1+t/M_\mathrm{s}^2)} \,.
\end{equation}
In a low-energy limit
 $M_\mathrm{s} \rightarrow \infty$,
 the string effects disappear as
 $V(s,t,u) \rightarrow 1$,
 and the string amplitudes become equal to the SM amplitudes
 (see Appendix \ref{sec:amplitude_and_width} for details).
In case of a finite value for $M_\mathrm{s}$,
 the form factor function is expanded by a sum over infinite $s$-channel poles,
\begin{equation}
	\label{eq:V-function_expanded}
 V(s,t,u)
  \simeq \sum_{n=1}^\infty
                \frac{1}{(n-1)!} \frac{1}{(M_\mathrm{s}^2)^{n-1}} \frac{1}{s-nM_\mathrm{s}^2}
                \prod_{J=0}^{n-1}
                \bigl(u+JM_\mathrm{s}^2\bigr) \,.
\end{equation}
This expansion is a good approximation,
 near each of $n$th pole
 $s \simeq n M_\mathrm{s}^2$.
These poles correspond to string excited states
 which have masses
 $M_n = \sqrt{n} M_\mathrm{s}$.
They are degenerated with different spins
 from $J = j_0$ to $J = j_0 + (n-1)$ for each $n$th pole,
 where $j_0$ is an original spin of initial states
 in the two-body scattering processes.
These string excited states are exchanged as virtual states in $s$-channel
 and are experimentally observed as resonances
 in these processes of the SM particles.
Signals by the string excited states at colliders
 were originally pointed out in Ref.\cite{Cullen:2000ef}.

If open strings have momenta and windings around
 in the direction of extra dimensions with general D$p$-branes ($p>3$),
 KK modes and winding modes of open strings
 appear in our four-dimensional space-time.
The masses and the ways to contribute to amplitudes of KK and winding modes
 depend on the details of the way of compactification.
Typical masses of quantized KK modes and winding modes are
 $M_n^\mathrm{KK}
  = n \bigl(V_{p-3}\bigr)^{-\frac{1}{p-3}}$
 and 
 $M_n^{\rm wind.}
  = n M_{\rm s}^2 \bigl(V_{p-3}\bigr)^{\frac{1}{p-3}}$,
 respectively.
Their masses approximate $n M_{\rm s}$
 and they are different from the masses of $n$th string excited states,
 $M_n = \sqrt{n} M_\mathrm{s}$.

Since closed string states interact only at one-loop level with the SM particles
 due to open-closed string duality,
 we do not consider these states.
Black holes may also be produced because of the low string scale,
 namely, the low fundamental gravitational scale.
However, since these produced black holes
 are expected to be evaporated by Hawking radiation,
 they do not give contributions to dijet events.
 We do not consider black holes also.

We concentrate on dijet events
 which are caused by parton two-body scattering processes at the LHC.
Since a target is the physics on resonances in the dijet invariant mass distribution,
 we may only discuss model independency of states
 which give $s$-channel poles in the scattering amplitudes.
If momenta in the direction of extra dimensions are conserved at interaction vertices,
 states of KK modes (winding modes) can appear only in pair
 and do not contribute to $s$-channel poles.
Momenta in the direction transverse to D-branes are non-conserved
 and momenta in the direction parallel to D-branes are conserved.
Therefore, even if D-branes contain extra dimensions,
 there are no $s$-channel exchanges of single KK modes (winding modes)
 in scattering processes of open strings on the D-branes.
In scattering processes of open strings connecting two different stacks of D-branes,
 however,
 there are $s$-channel exchanges of single KK modes (winding modes)
 due to non-conservation of momenta.

\begin{figure}
	\centering
	\begin{overpic}[width=16cm]{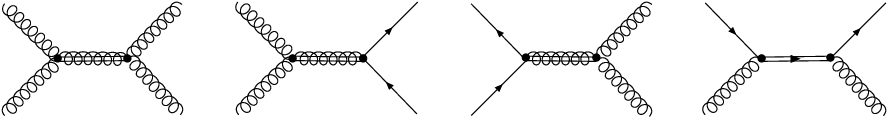}
		\put(45,15){$g^*$}
		\put(165,15){$g^*$}
		\put(285,15){$g^*$}
		\put(405,15){$q^*$}
	\end{overpic}
	\caption{The model independent two-parton scattering processes
	                with exchanges of string excited states of gluons and quarks,
	                $g^*$ and $q^*$, respectively.}
	\label{fig:2-parton_scattering_process_model_independent}
\vspace{8mm}
	\centering
	\begin{overpic}[width=4cm]{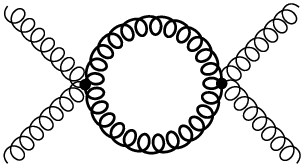}
		\put(50,-5){$g_\mathrm{KK}$}
		\put(50,62){$g_\mathrm{KK}$}
	\end{overpic}
	\caption{The $gg\rightarrow gg$ process with the exchange
	                 of a pair of KK gluons $g_\mathrm{KK}$.}
	\label{fig:KK_gluon_pair_exchange}
\vspace{6mm}
	\centering
	\begin{overpic}[width=8cm]{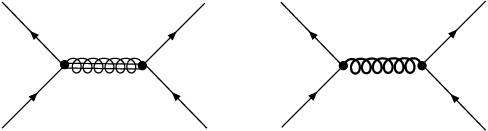}
		\put(45,15){$g^*$}
		\put(170,15){$g_\mathrm{KK}$}
	\end{overpic}
	\caption{The model-dependent processes
	               with the exchange of a KK gluon $g_\mathrm{KK}$.}
	\label{fig:2-parton_scattering_process_model_dependent}
\end{figure}

The processes of
 $gg \rightarrow gg$,
 $gg \rightarrow q\bar{q}$
 and $q\bar{q} \rightarrow gg$
 are model independent.
In these processes,
 only string excited states of gluons can contribute to $s$-channel poles
 as shown in Fig.\ref{fig:2-parton_scattering_process_model_independent},
 and KK modes (winding modes) of gluons can appear only in pair
 as shown in Fig.\ref{fig:KK_gluon_pair_exchange}.
The process of $qg\rightarrow qg$ is also model independent
 with string excited states of quarks in $s$-channel
 as shown in Fig.\ref{fig:2-parton_scattering_process_model_independent},
 because momenta on the intersection plane
 where open strings of chiral fermions are localized are conserved.
On the other hand, the process of $q\bar{q}\rightarrow q\bar{q}$ is model dependent,
 since both of KK modes (winding modes) of gluons
 and gluon excitations are exchanged in $s$-channel
 as shown in Fig.\ref{fig:2-parton_scattering_process_model_dependent}.
Fortunately, cross sections of this process are suppressed
 by the effect of parton distribution functions (PDFs) at the LHC as a proton-proton collider.
We concentrate on the above model-independent processes at the LHC,
 and discuss leading string signals which do not depend on the details of model buildings.

In Section \ref{sec:1st_resonance},
 the appearance of the resonance in the dijet invariant mass distribution
 due to first string excited states is reviewed.
In Section \ref{sec:angular_analysis},
 it is shown that spin degeneracy at the resonance can be observed
 by analyzing the dijet angular distribution.
In Section \ref{sec:2nd_resonance},
 it is shown that the second resonance can be observed
 at rather close to the first resonance in the dijet invariant mass distribution.
In Section \ref{sec:conclusion},
 we give a summary of our results and some discussions.
Some detailed properties of amplitudes and widths of first and second string excited states
 are summarized in Appendix \ref{sec:amplitude_and_width}.

\section{The first string resonances}
	\label{sec:1st_resonance}

	\hspace{5mm}
Spin-averaged squared amplitudes
 of the model-independent parton two-body scattering processes
 with exchanges of the first string excited states are calculated
 in Ref.\cite{Anchordoqui:2009mm,Anchordoqui:2008di,Anchordoqui:2008hi}.
\begin{equation}
	\begin{split}
	\label{eq:1st_excited_squared_amplitude_gg-gg}
 \bigl| \mathcal{M}_\mathrm{1st} ( gg \rightarrow gg ) \bigr|^2
  = & \frac{8}{N^2}
  	  \frac{g^4}{M_\mathrm{s}^4}
	  \Biggl\{ 
         	\frac{(N^2-4)^2}{4(N^2-1)}
          	\Biggl[
              		\frac{ M_\mathrm{s}^8 }
			         { ( \hat{s} - M_\mathrm{s}^2 )^2
			         + \bigl( M_\mathrm{s} \Gamma_{g^*,\mathrm{1st}}^{J=0} \bigr)^2 }
           	    + \frac{ \hat{u}^4 + \hat{t}^4 }
              		         { ( \hat{s} - M_\mathrm{s}^2 )^2
              		         + \bigl( M_\mathrm{s} \Gamma_{g^*,\mathrm{1st}}^{J=2} \bigr)^2 }
		\Biggr] \\
  & \hspace{30mm}
	    + \Biggl[
	    		\frac{ M_\mathrm{s}^8 }
			         { ( \hat{s} - M_\mathrm{s}^2 )^2
			         + \bigl( M_\mathrm{s} \Gamma_{C^*,\mathrm{1st}}^{J=0} \bigr)^2 }
		    + \frac{ \hat{u}^4 + \hat{t}^4 }
		    		  { ( \hat{s} - M_\mathrm{s}^2 )^2
				  + \bigl( M_\mathrm{s} \Gamma_{C^*,\mathrm{1st}}^{J=2} \bigr )^2}
		\Biggr]
	\Biggr\} \,,
	\end{split}
\end{equation}
\begin{equation}
	\label{eq:1st_excited_squared_amplitude_gg-qq}
 \bigl| \mathcal{M}_\mathrm{1st} ( gg \rightarrow q\bar{q} ) \bigr|^2
  = \frac{2}{N(N^2-1)}
     N_f
     \frac{g^4}{M_\mathrm{s}^4}
     \Biggl[\frac{N^2-4}{2}
		\frac{\hat{u}\hat{t} (\hat{u}^2 + \hat{t}^2)}
		         { ( \hat{s} - M_\mathrm{s}^2 )^2
		         + \bigl( M_\mathrm{s} \Gamma_{g^*,\mathrm{1st}}^{J=2} \bigr)^2 }
	    + \frac{ \hat{u}\hat{t} ( \hat{u}^2 + \hat{t}^2 )}
	    		  { ( \hat{s} - M_\mathrm{s}^2 )^2
			  + \bigl( M_\mathrm{s} \Gamma_{C^*,\mathrm{1st}}^{J=2} \bigr)^2 }
      \Biggr] \,,
\end{equation}
\begin{equation}
	\label{eq:1st_excited_squared_amplitude_qq-gg}
 \bigl| \mathcal{M}_\mathrm{1st} ( q\bar{q} \rightarrow gg ) \bigr|^2
 = \frac{2(N^2-1)}{N^3}
    \frac{g^4}{M_\mathrm{s}^4}
    \Biggl[
    		\frac{N^2-4}{2}
		\frac{ \hat{u}\hat{t} ( \hat{u}^2 + \hat{t}^2 ) }
		         { ( \hat{s} - M_\mathrm{s}^2 )^2
		         + \bigl( M_\mathrm{s} \Gamma_{g^*,\mathrm{1st}}^{J=2} \bigr)^2 }
	    + \frac{ \hat{u}\hat{t} ( \hat{u}^2 + \hat{t}^2 ) }
	                { ( \hat{s} - M_\mathrm{s}^2 )^2
	                + \bigl( M_\mathrm{s} \Gamma_{C^*,\mathrm{1st}}^{J=2} \bigr)^2 }
    \Biggr] \,,
\end{equation}
\begin{equation}
	\begin{split}
	\label{eq:1st_excited_squared_amplitude_qg_t-channel}
 \bigl| \mathcal{M}_\mathrm{1st} ( qg \rightarrow qg ) \bigr|^2
 = & 
 \bigl| \mathcal{M}_\mathrm{1st} ( \bar{q}g \rightarrow \bar{q}g ) \bigr|^2 \\
 = & \frac{N^2-1}{2N^2}
 	 \frac{g^4}{M_\mathrm{s}^2}
	 \Biggl[
	 	\frac{ M_\mathrm{s}^4 (-\hat{u}) }
		         { ( \hat{s} - M_\mathrm{s}^2 )^2
		         + \bigl( M_\mathrm{s} \Gamma_{q^*,\mathrm{1st}}^{J=1/2} \bigr)^2 }
	    + \frac{ (-\hat{u})^3 }
	                { ( \hat{s} - M_\mathrm{s}^2 )^2
	                + \bigl( M_\mathrm{s} \Gamma_{q^*,\mathrm{1st}}^{J=3/2} \bigr)^2 }
	 \Biggr] \,,
	 \hspace{10mm}
	\end{split}
\end{equation}
\begin{equation}
	\begin{split}
 	\label{eq:1st_excited_squared_amplitude_qg_u-channel}
 \bigl| \mathcal{M}_\mathrm{1st} ( qg \rightarrow gq) \bigr|^2
 = & 
 \bigl| \mathcal{M}_\mathrm{1st} ( \bar{q}g \rightarrow g\bar{q} ) \bigr|^2 \\
 = & \frac{N^2-1}{2N^2}
 	 \frac{g^4}{M_\mathrm{s}^2}
	 \Biggl[
	 	\frac{ M_\mathrm{s}^4 (-\hat{t}) }
		         { ( \hat{s} - M_\mathrm{s}^2 )^2
		         + \bigl( M_\mathrm{s} \Gamma_{q^*,\mathrm{1st}}^{J=1/2} \bigr)^2 }
	    + \frac{ (-\hat{t})^3 }
	                { ( \hat{s} - M_\mathrm{s}^2 )^2
	                + \bigl( M_\mathrm{s} \Gamma_{q^*,\mathrm{1st}}^{J=3/2} \bigr)^2}
	  \Biggr] \,,
	  \hspace{10mm}
	 \end{split}
\end{equation}
where $N = 3$, $N_f = 6$ and
 $\Gamma_{g^*,C^*,q^*}^{J}$ are total decay widths of the first excited states
 of gluons, the U$(1)_\mathrm{color}$ gauge boson and quarks with spin $J$, respectively,
\begin{equation}
	\label{eq:1st_excited_width_g}
 \Gamma_{g^*,\mathrm{1st}}^{J=0}
 = \frac{g^2}{4\pi}
    M_\mathrm{s}
    \frac{N}{4} \,,
    \hspace{6mm}
 \Gamma_{g^*,\mathrm{1st}}^{J=2}
 = \frac{g^2}{4\pi}
    M_\mathrm{s}
    \biggl( \frac{N}{10} + \frac{N_f}{40} \biggr) \,,
\end{equation}
\begin{equation}
	\label{eq:1st_excited_width_C}
 \Gamma_{C^*,\mathrm{1st}}^{J=0}
 = \frac{g^2}{4\pi}
    M_\mathrm{s}
    \frac{N}{2} \,,
    \hspace{6mm}
 \Gamma_{C^*,\mathrm{1st}}^{J=2}
 = \frac{g^2}{4\pi}
    M_\mathrm{s}
    \biggl( \frac{N}{5} + \frac{N_f}{40} \biggr) \,,
\end{equation}
\begin{equation}
	\label{eq:1st_excited_width_q}
 \Gamma_{q^*,\mathrm{1st}}^{J=1/2}
 = \frac{g^2}{4\pi}
    M_\mathrm{s}
    \frac{N}{8} \,,
    \hspace{6mm}
 \Gamma_{q^*,\mathrm{1st}}^{J=3/2}
 = \frac{g^2}{4\pi}
    M_\mathrm{s}
    \frac{N}{16} \,,
    \hspace{15.5mm}
\end{equation}
where $g$ is the gauge coupling constant of strong interaction\footnote{The value of $\Gamma^{J=3/2}_{q^*}$
                   is different by factor $1/2$ from the one in Ref.\cite{Anchordoqui:2008hi}.
                  See Appendix \ref{sec:amplitude_and_width} for details.}.
Here, $\hat{s}$, $\hat{t}$ and $\hat{u}$ are the Mandelstam variables of partons.
The squared amplitudes
 of eqs.(\ref{eq:1st_excited_squared_amplitude_gg-gg})-(\ref{eq:1st_excited_squared_amplitude_qg_u-channel})
 are ``minimal",
 because we do not consider contributions in $\hat{t}$- and $\hat{u}$-channels,
 which should be a good approximation for any low-scale string models
 at least near the poles of first string excited states,
 $\hat{s} \simeq M_\mathrm{s}^2$.
Note that all these first string excited states are degenerated in mass at tree level,
 though they have different spins and decay widths.

Two partons in final states are hadronized
 and give dijet events at the LHC.
The dijet invariant mass $M_{jj}$ is an important observable,
 since a peak or resonance at the mass of string excited states in the $M_{jj}$ distribution
 may be observed.
To make a prediction to the dijet invariant mass distribution,
 we have to do computer simulations for hadronization and detector simulation.
There have been few such works on low-scale string models.
We perform Monte Carlo simulations by using
 $\mathtt{CalcHEP}$ \cite{Pukhov:1999gg} for event generation,
 $\mathtt{PYTHIA\,8}$ \cite{Sjostrand:2007gs} for hadronization,
 $\mathtt{Delphes\,1.9}$ \cite{Ovyn:2009tx} for detector simulation
 using its default detector card for ATLAS,
 and $\mathtt{ROOT}$ \cite{Brun:1997pa} for analysis of event samples.
The details of event generations with string resonances
 are found in the web page of \cite{event_generation_with_string}.
In this paper, we do not consider interference effects between the SM processes
 and the processes with string excited states,
 because the latter completely dominates at the resonances.

\begin{figure}[t]
	\centering
	\includegraphics[width=12cm]
	{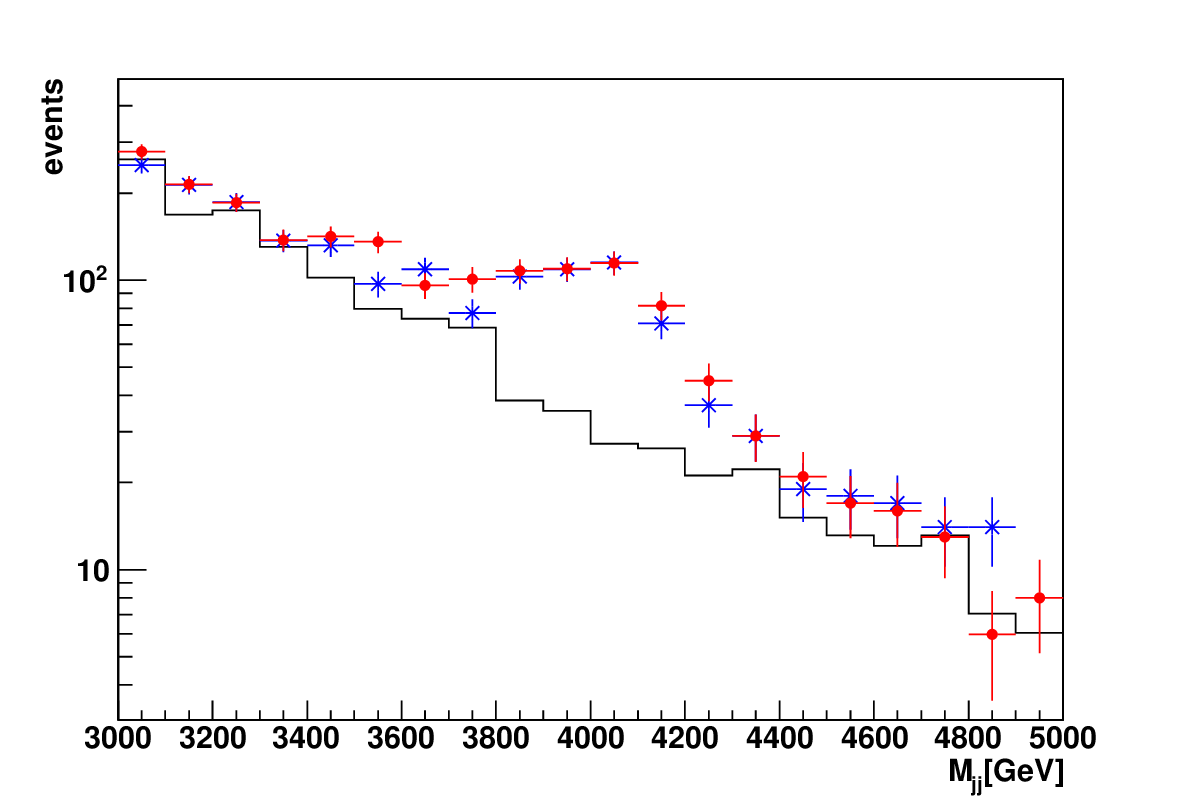}
	\caption{The dijet invariant mass distribution for $M_\mathrm{s} = 4\,\mathrm{TeV}$
	                 with $1.4\,\mathrm{fb}^{-1}$ of integrated luminosity
	                 at $\sqrt{s} = 14\,\mathrm{TeV}$.
	                The red distribution includes all subprocesses
	                 of eqs.(\ref{eq:1st_excited_squared_amplitude_gg-gg})-(\ref{eq:1st_excited_squared_amplitude_qg_u-channel}),
	                 while the blue distribution includes only dominant subprocesses,
	                 $qg\rightarrow qg$,
	                 of eqs.(\ref{eq:1st_excited_squared_amplitude_qg_t-channel})-(\ref{eq:1st_excited_squared_amplitude_qg_u-channel}).
	                The black histogram shows the SM background.}
	\label{fig:1st_excited_dijet_mass}
\end{figure}

In the rest of paper, we chose a value
 $M_\mathrm{s} = 4\,\mathrm{TeV}$ as a reference,
 though the CMS experiment gives a bound
 $M_\mathrm{s} > 4\,\mathrm{TeV}$ \cite{Chatrchyan:2011ns}.
Fig.\ref{fig:1st_excited_dijet_mass} shows the result of simulations
 of the dijet invariant mass distribution
 for $M_\mathrm{s} = 4\,\mathrm{TeV}$
 and $1.4\,\mathrm{fb}^{-1}$ of integrated luminosity
 at $\sqrt{s} = 14\,\mathrm{TeV}$.
We apply cuts
 $p_{\mathrm{T},j_1} > 330\,\mathrm{GeV}$,
 $|y_1|$, $|y_2| < 2.5$
 and $|y_1-y_2| < 1.3$,
 where $p_{\mathrm{T},j_1}$ is a largest transverse momentum of jets,
 and $y_1$ and $y_2$ are pseudo-rapidities of first and second jets
 which have primary and secondary large $p_\mathrm{T}$s.
We see that contributions from the processes of $qg \rightarrow qg$
 are dominant near the resonance
 at $M_\mathrm{s} = 4\,\mathrm{TeV}$,
 because of the dominance of quarks in the PDF at high energies.
We consider string contributions to $qg \rightarrow qg$ processes only in the following.

\section{Angular analysis}
	\label{sec:angular_analysis}

	\hspace{5mm}
The dijet angular distribution exhibits
 a distinct property in low-scale string models
 because of spin degeneracy of string excited states.
We consider only the process of $qg\rightarrow qg$,
 because the cross section of the process is dominant
 near the resonance in the dijet invariant mass distribution.
In the process, two first string excited states of quarks with $J=1/2$ and $J=3/2$
 are exchanged in $s$-channel,
 and they are degenerated in mass.
If we can experimentally confirm the degeneracy of states,
 it is a signature of low-scale string models.

We analyze the $\chi$-distribution on the resonance,
 which was used by ATLAS experiment to search ``new physics" beyond the SM
 \cite{Aad:2011aj} (see Ref.\cite{Boelaert:2009jm} for details).
The quality $\chi$ is defined as
\begin{equation}
	\label{eq:chi}
 \chi
 = \exp(y_1-y_2)
 = \frac{1+\cos\theta_*}{1-\cos\theta_*} \,,
\end{equation}
where $y_1$ and $y_2$ are pseudo-rapidities of two jets
 and $\theta_*$ is a scattering angle in the parton center-of-mass frame.
For any ``new physics" which gives resonances by heavy states,
 the number of events for small $\chi \sim 1$ is enhanced
 because of enhancement of scatterings with large angles $\cos\theta_*\sim0$.
In the SM, the $\chi$-distribution is flat
 since $t$- and $u$-channel exchanges are dominant.

A formula of the $\chi$-distribution is derived in the following way.
A cross section of dijet events for a two-parton scattering process
 is described as
\begin{equation}
	\begin{split}
	\label{eq:2-parton_scattering_process_cross_section_at_LHC}
 \sigma \bigl( \mathrm{p}_1(P_1), \mathrm{p}_2(P_2)
 			     \rightarrow & j_1(p_1), j_2(p_2), X \bigr) \\
 = & \sum_i
 	 \sum_j
	 \int^1_0 dx_1
	 \int^1_0 dx_2
	 	f_i(x_1)
		f_j(x_2)
		\sigma \bigl( i(p_i), j(p_j) \rightarrow k(p_k), \ell(p_\ell) \bigr) \,,
	\end{split}
\end{equation}
where $\mathrm{p}_1$, $\mathrm{p}_2$ and $X$ denote
 incoming protons and QCD remnants
 and $j_1$, $j_2$ denote observed jets,
 and $f_i(x_1)$, $f_j(x_2)$ are PDFs of protons
 for initial partons $i$, $j$ with the momentum fractions
\begin{equation}
	\label{eq:momentum_fraction}
 x_1
 = \frac{p_i}{P_1} \,, \hspace{6mm}
 x_2
 = \frac{p_j}{P_2} \,,
\end{equation}
 respectively.
A cross section of a specific two-parton scattering process
 $ij \rightarrow k\ell$ can be recast into
\begin{equation}
	\label{eq:2-parton_scattering_process_cross_section}
 \sigma ( ij \rightarrow k\ell )
 = \int^1_0 dx_1
    \int^1_0 dx_2
    		f_i(x_1)
		f_j(x_2)
		\int d\hat{t}
			\frac{d\sigma ( ij \rightarrow k\ell )}
				  {d\hat{t}} \,,
\end{equation}
where the differential cross section is simply described
 by a spin-averaged squared amplitude $| \mathcal{M}(ij \rightarrow k\ell) |^2$ as
\begin{equation}
	\label{eq:2-parton_scattering_process_differential_cross_section}
 \frac{d\sigma ( ij \rightarrow k\ell )}
          {d\hat{t}}
 = \frac{\bigl| \mathcal{M} ( ij \rightarrow k\ell ) \bigr|^2}
             {16\pi \hat{s}^2} \,.
\end{equation}
The formula of the $\chi$-distribution is obtained
 by performing a change of integrating variables
 in eq.(\ref{eq:2-parton_scattering_process_cross_section}).

In the parton center-of-mass frame,
 rapidities of partons $k$ and $\ell$ are opposite in sign:
 $y \equiv y_k^* = -y_\ell^*$,
 since these partons are produced back-to-back in this frame.
When we define a boost velocity
 from the parton center-of-mass frame to the proton center-of-mass frame,
 $\beta \equiv \tanh Y$,
 pseudo-rapidities of observed jets, $y_1$ and $y_2$, are written as
\begin{equation}
	\label{eq:rapidity_dijet}
 y_1 =   y + Y \,, \hspace{6mm}
 y_2 = - y + Y \,.
\end{equation}
The quantities
\begin{equation}
	\label{eq:rapidity_parton_and_boost}
 y  = \frac{1}{2} (y_1 - y_2) \,, \hspace{6mm}
 Y = \frac{1}{2} (y_1 + y_2) \,,
\end{equation}
 are independent observables.
The necessary kinetic variables
 in eq.(\ref{eq:2-parton_scattering_process_cross_section})
 are described by $\chi = \exp(2y)$,
                                $Y$,
                                and the dijet invariant mass $M = \sqrt{\hat{s}}$ as
\begin{equation}
	\label{eq:t-hat_and_u-hat}
 \hat{t}  = - \frac{M^2}{1 + \chi} \,, \hspace{6mm}
 \hat{u} = - \frac{M^2 \chi}{1 + \chi} \,,
\end{equation}
\begin{equation}
	\label{eq:momentum_fraction_2}
 x_1 = \sqrt{\frac{M^2}{s}} e^Y \,, \hspace{6mm}
 x_2 = \sqrt{\frac{M^2}{s}} e^{-Y} \,,
\end{equation}
 where $\sqrt{s}$ is the proton center-of-mass energy.
Then we have
\begin{equation}
	\label{eq:2-parton_scattering_process_cross_section_2}
 \sigma ( ij \rightarrow k\ell )
 = \int dM^2
    \int dY
    	x_1 f_i(x_1)
	x_2 f_j(x_2)
	\int d\chi
		\frac{1}{(1+\chi)^2}
		\frac{d\sigma ( ij \rightarrow k\ell )}
		         {d\hat{t}} \,.
\end{equation} 
Inserting the squared amplitude
 of eqs.(\ref{eq:1st_excited_squared_amplitude_qg_t-channel})-(\ref{eq:1st_excited_squared_amplitude_qg_u-channel})
 into eq.(\ref{eq:2-parton_scattering_process_cross_section_2}),
 a prediction to the $\chi$-distribution from low-scale string models is obtained as
\begin{equation}
	\label{eq:qg_scattering_process_cross_section_chi_dependence}
 \frac{d\sigma ( qg \rightarrow qg )}{d\chi}
 = \frac{1}{(1+\chi)^2}
    \biggl(
    		C_{1/2}
	    + C_{3/2} \frac{1 + \chi^3}{(1 + \chi)^3} \,,
    \biggr)
\end{equation}
 where $C_{1/2}$ and $C_{3/2}$ are constants.
The term proportional to $C_{1/2}$
 represents $\chi$-dependence of events with exchanges of $J=1/2$ states
 and the term proportional to $C_{3/2}$
 represents that of $J=3/2$ states.
The factor $1/(1+\chi)^2$
 in eq.(\ref{eq:qg_scattering_process_cross_section_chi_dependence})
 is a kinematical factor.

Event samples for the SM only
 and for the SM with first string excited states
 in the subprocesses of $qg\rightarrow qg$ are generated.
The numbers of generated events
 are corresponding to $1.8\,\mathrm{fb}^{-1}$
 and $18.7\,\mathrm{fb}^{-1}$ of integrated luminosity
 at $\sqrt{s} = 14\,\mathrm{TeV}$.
 Kinematical cuts imposed in this analysis are
\begin{equation}
	\label{eq:analysis_cut}
	\begin{split}
 & \hspace{21mm}
 p_{\mathrm{T},j_1} > 350\,\mathrm{GeV} \,, \\
 & M_\mathrm{s} - 250\,\mathrm{GeV}
     < M_{jj}
     < M_\mathrm{s} + 250\,\mathrm{GeV} \,, \\
 & \hspace{6mm}
 |y_1 - y_2| < 2.3 \,, \hspace{6mm}
 |y_1 + y_2| < 2.0 \,. 
	\end{split}
\end{equation}
The same analysis is applied to both event samples
 for the SM and for the SM with string excited states
 in the subprocesses of $qg \rightarrow qg$,
 and we obtain signal event samples
 by subtracting the first one from the second one.

\begin{figure}
	\centering
	\includegraphics[width=12cm]{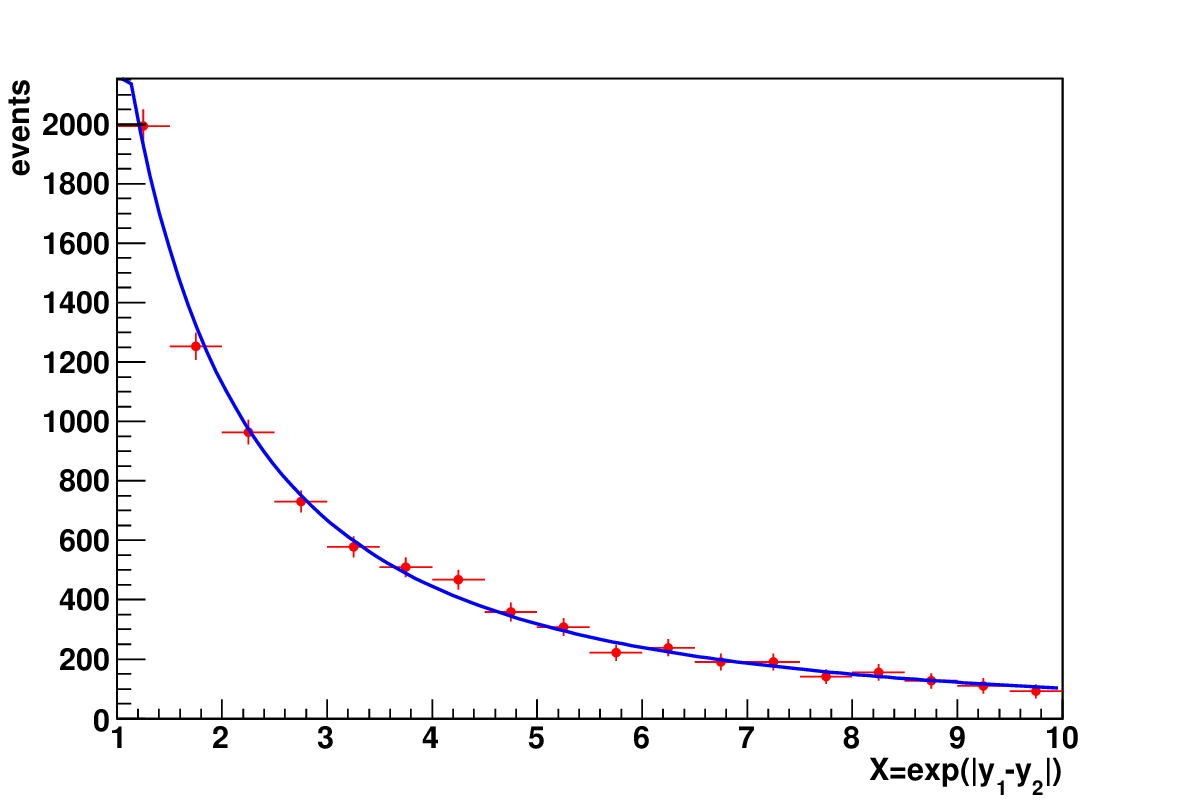}
	\caption{The $\chi$-distribution for $M_\mathrm{s} = 4\,\mathrm{TeV}$
	                 with $18.7\,\mathrm{fb}^{-1}$ of integrated luminosity
	                 at $\sqrt{s} = 14\,\mathrm{TeV}$. 
	                The blue line is a fit with both $J=1/2$ and $3/2$ states.}
	\label{fig:spin1-2_and_spin3-2_chi}
\end{figure}

\begin{figure}
	\centering
	\includegraphics[width=12cm]{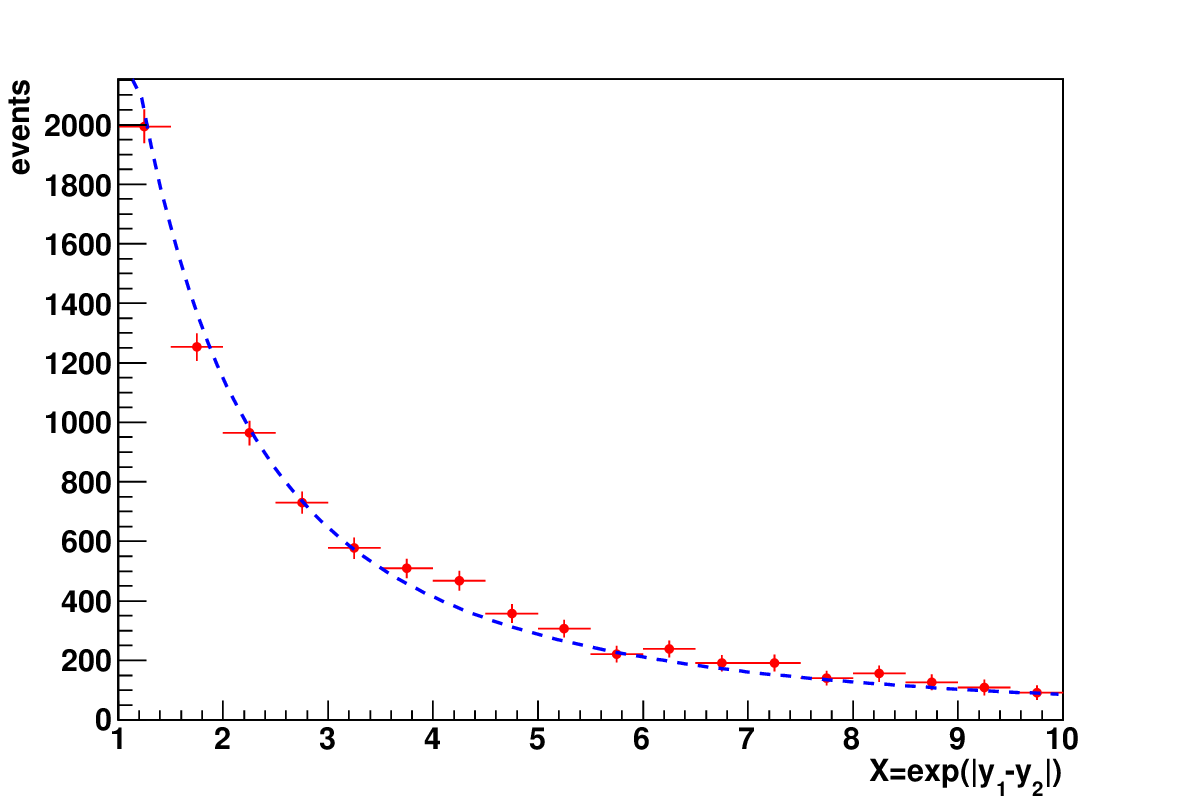}
	\caption{The $\chi$-distribution for $M_\mathrm{s} = 4\,\mathrm{TeV}$
	                 with $18.7\,\mathrm{fb}^{-1}$ of integrated luminosity
	                 at $\sqrt{s} = 14\,\mathrm{TeV}$. 
	                The blue short dashed line is a fit with only a $J=1/2$ state.}
	\label{fig:spin1-2_only_chi}
\end{figure}

\begin{figure}
	\centering
	\includegraphics[width=12cm]{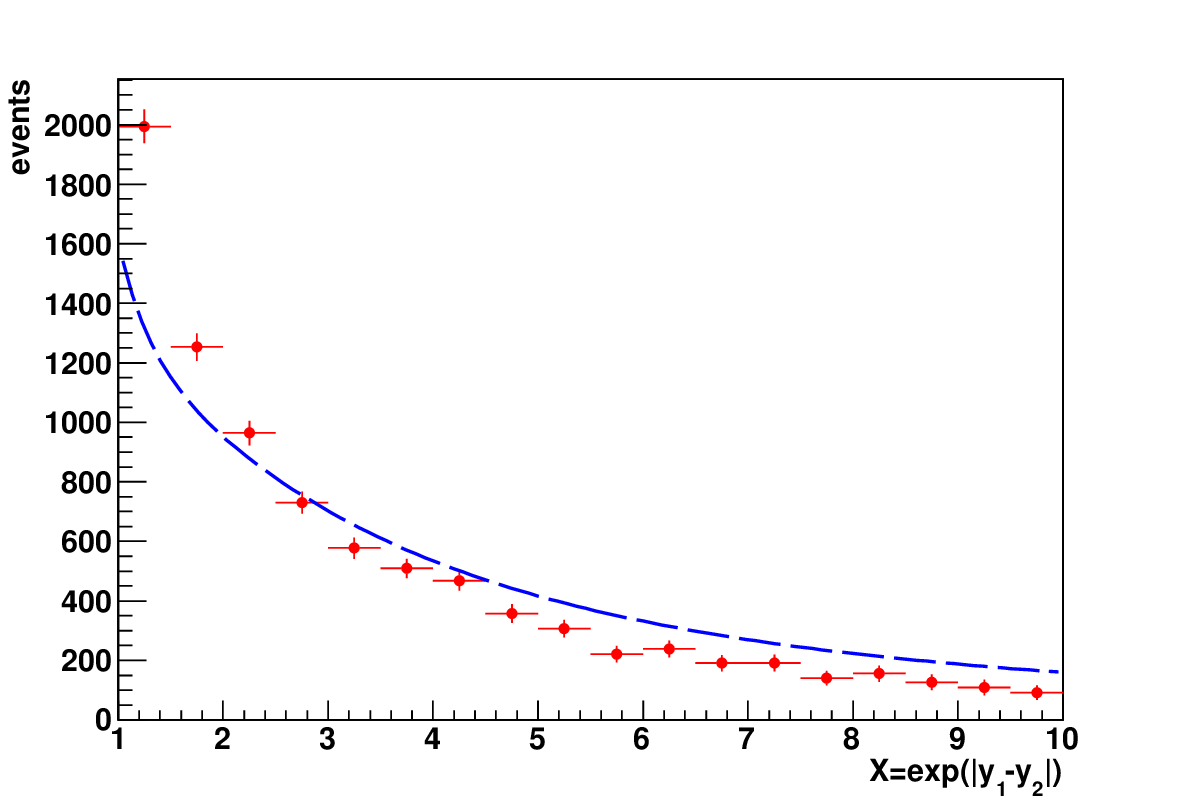}
	\caption{The $\chi$-distribution for $M_\mathrm{s} = 4\,\mathrm{TeV}$
	                 with $18.7\,\mathrm{fb}^{-1}$ of integrated luminosity
	                 at $\sqrt{s} = 14\,\mathrm{TeV}$. 
	                The blue long dashed line is a fit with only a $J=3/2$ state.}
	\label{fig:spin3-2_only_chi}
\end{figure}

We fit the $\chi$-distribution of signal events
 with functions of $\chi$
 in eq.(\ref{eq:qg_scattering_process_cross_section_chi_dependence})
 in three cases of $C_{1/2}\neq0$ and $C_{3/2}\neq0$,
                               $C_{1/2}\neq0$ and $C_{3/2}=0$,
                               and $C_{1/2}=0$ and $C_{3/2}\neq0$.
These three cases correspond to assumptions
 with both $J=1/2$ and $3/2$ states, only a $J=1/2$ state
 and only a $J=3/2$ state, respectively.
Assuming the state with only $J=1/2$, for example,
 is corresponding to consider new quark-like particles of some other ``new physics".
The results of these fits are shown
 in Figs.\ref{fig:spin1-2_and_spin3-2_chi}-\ref{fig:spin3-2_only_chi}.

We clearly see in Fig.\ref{fig:spin3-2_only_chi} that the fit with only $J=3/2$ is not good.
The difference of goodness-of-fit
 between the fit with both $J=1/2$ and $3/2$ and the one with only $J=1/2$,
 however, are visually not apparent.
The $p$-values of these fits are shown in Table \ref{tab:p-value}.

\begin{table}[h]
	\begin{center}
	\begin{tabular}{|c|c|c|c|}\hline
	                                              & $J=1/2$ and $3/2$
	 & $J=1/2$ only             & $J=3/2$ only                                   \\ \hline
	$9.8\,\mathrm{fb}^{-1}$   & $0.003734\,(2.20)$
	 & $0.001187\,(2.37)$ & $1.248\times10^{-31}\,(11.28)$ \\
	$18.7\,\mathrm{fb}^{-1}$ & $0.7471\,(0.747)$
	 & $0.04058\,(1.67)$   & $0.\,(16.84)$                                   \\ \hline
	\end{tabular}
	\end{center}
	\caption{$p$-values (reduced $\chi^2$ values) of the fits}
	\label{tab:p-value}
\end{table}
The $p$-value, in brief,
 is a probability that a hypothesis is excluded {\it by error}
 in spite of that the hypothesis is correct \cite{Cowan:2008zza}.
Namely, if the $p$-value is small,
 we can exclude the hypothesis.
With $9.8\,\mathrm{fb}^{-1}$ of integrated luminosity,
 $p$-values of the fits with $J=1/2$ and $3/2$ and with only $J=1/2$
 are of the same order and very small.
 It is difficult to distinguish them statistically.
However, with $18.7\,\mathrm{fb}^{-1}$ of integrated luminosity,
 the $p$-value of the fit with $J=1/2$ and $3/2$ is much larger
 than that of the fit with only $J=1/2$.
Since a significance level of $p$-values to exclude a hypothesis
 is usually taken as $0.05$,
 the fit with $J=1/2$ and $3/2$ in which the $p$-value is $0.75$
 much higher than $0.05$ can be said to be good,
 and the fit with only $J=1/2$ in which the $p$-value is lower than $0.05$
 can be excluded.
If we look at the $\chi$-distribution of Fig.\ref{fig:spin1-2_only_chi} closer,
 we see that
 the fit in the region $\chi\gtrsim3.8$ is systematically inconsistent.

We conclude that spin degeneracy at the resonance
 in the dijet invariant mass distribution by string excited states
 can be experimentally confirmed.
This can be a signature of low-scale string models.

\section{The second string resonances}
	\label{sec:2nd_resonance}

	\hspace{5mm}
Another distinct property of low-scale string models
 is the appearance of the second resonance
 in the dijet invariant mass distribution at a specific value of $M_{jj}$.
Second string excited states have characteristic masses of
 $M_\mathrm{2nd} = \sqrt{2} M_\mathrm{s}$,
 while second KK modes of the SM particles,
 for example, have typical masses of
 $M_\mathrm{2nd}^\mathrm{KK} \sim 2M$
 where $M$ is the mass of first KK modes.
If a low-scale string model is realized,
 the second resonance should exist
 at rather close to the first resonance in the dijet invariant mass distribution.

We calculate spin-averaged squared amplitudes of the dominant processes,
 $qg \rightarrow qg$ and $qg\rightarrow gq$,
 with exchanges of second string excited states.
\begin{equation}
	\begin{split}
	\label{eq:2nd_excited_squared_amplitude_t-channel}
 & \bigl| \mathcal{M}_\mathrm{2nd} ( qg \rightarrow qg ) \bigr|^2
 = \bigl| \mathcal{M}_\mathrm{2nd} ( \bar{q}g \rightarrow \bar{q}g ) \bigr|^2 \\
 = & \frac{2(N^2-1)}{N^2}
 	 \Biggl\{
		 \frac{g^4}{2M_\mathrm{s}^2}
		 \Biggl[
		 	\hspace{2mm}
		 	\frac{1}{9}
			\frac{ M_\mathrm{s}^4 ( -\hat{u} ) }
			        { \bigl( \hat{s} - 2M_\mathrm{s}^2 \bigr)^2
			       + \bigl( \sqrt{2}M_\mathrm{s} \Gamma^{J=1/2}_{q^*,\mathrm{2nd}} \bigr)^2 }
			\hspace{1mm}
		   + \hspace{1mm}
			\frac{1}{9}
			\frac{ ( -\hat{u} ) ( 3\hat{t} + \hat{s} )^2 }
			         { \bigl( \hat{s} - 2M_\mathrm{s}^2 \bigr)^2
			        + \bigl( \sqrt{2}M_\mathrm{s} \Gamma^{J=3/2}_{q^*,\mathrm{2nd}} \bigr)^2 }
		\Biggr] \\
	    & \hspace{17mm}
	    + \frac{g^4}{8M_\mathrm{s}^6}
	       \Biggl[
	    		\frac{9}{25}
			\frac{ M_\mathrm{s}^4 ( -\hat{u} )^3}
			         { \bigl( \hat{s} - 2M_\mathrm{s}^2 \bigr)^2
			        + \bigl( \sqrt{2}M_\mathrm{s} \Gamma^{J=3/2}_{q^*,\mathrm{2nd}} \bigr)^2 }
		    + \frac{1}{25}
		       \frac{ ( -\hat{u} )^3 ( 5\hat{t} + \hat{s} )^2 }
		                { \bigl( \hat{s} - 2M_\mathrm{s}^2 \bigr)^2
		               + \bigl( \sqrt{2}M_\mathrm{s} \Gamma^{J=5/2}_{q^*,\mathrm{2nd}} \bigr)^2 }
		 \Biggr]
	\Biggr\} \,,
	\end{split}
	\end{equation}
\begin{equation}
	\begin{split}
	\label{eq:2nd_excited_squared_amplitude_u-channel}
 & \bigl| \mathcal{M}_\mathrm{2nd} ( qg \rightarrow gq ) \bigr|^2
 = \bigl| \mathcal{M}_\mathrm{2nd} ( \bar{q}g \rightarrow g\bar{q} ) \bigr|^2 \\
 = & \frac{2(N^2-1)}{N^2}
 	 \Biggl\{
	 	\frac{g^4}{2M_\mathrm{s}^2}
		\Biggl[
			\hspace{2mm}
			\frac{1}{9}
			\frac{ M_\mathrm{s}^4 ( -\hat{t} ) }
			         { \bigl( \hat{s} - 2M_\mathrm{s}^2 \bigr)^2
			        + \bigl( \sqrt{2}M_\mathrm{s} \Gamma^{J=1/2}_{q^*,\mathrm{2nd}} \bigr)^2 }
			\hspace{1mm}
		    + \hspace{1mm}
		       \frac{1}{9}
		       \frac{ ( -\hat{t} ) ( 3\hat{u} + \hat{s} )^2 }
		                { \bigl( \hat{s} - 2M_\mathrm{s}^2 \bigr)^2
		               + \bigl( \sqrt{2}M_\mathrm{s} \Gamma^{J=3/2}_{q^*,\mathrm{2nd}} \bigr)^2 }
		 \Biggr] \\
	    & \hspace{17mm}
	    + \frac{g^4}{8M_\mathrm{s}^6}
 		\Biggl[
			\frac{9}{25}
			\frac{ M_\mathrm{s}^4 ( -\hat{t} )^3}
			         { \bigl( \hat{s} - 2M_\mathrm{s}^2 \bigr)^2
			        + \bigl( \sqrt{2}M_\mathrm{s} \Gamma^{J=3/2}_{q^*,\mathrm{2nd}} \bigr)^2 }
		    + \frac{1}{25}
		       \frac{ ( -\hat{t} )^3 ( 5\hat{u} + \hat{s} )^2 }
		                { \bigl( \hat{s} - 2M_\mathrm{s}^2 \bigr)^2
		               + \bigl( \sqrt{2}M_\mathrm{s} \Gamma^{J=5/2}_{q^*,\mathrm{2nd}} \bigr)^2 }
 		\Biggr]
	\Biggr\} \,.
	\end{split}
\end{equation}
Total widths of the second excited states of quarks with spin $J$,
 $\Gamma^J_{q^*,\mathrm{2nd}}$, are given by
\begin{equation}
	\label{eq:2nd_excited_width}
 \Gamma^{J=1/2}_{q^*,\mathrm{2nd}}
 = \frac{g^2}{4\pi}
    \sqrt{2}M_\mathrm{s}
    \frac{N}{24} \,,
    \hspace{6mm}
 \Gamma^{J=3/2}_{q^*,\mathrm{2nd}}
 = \frac{g^2}{4\pi}
    \sqrt{2}M_\mathrm{s}
    \frac{19N}{240} \,,
    \hspace{6mm}
 \Gamma^{J=5/2}_{q^*,\mathrm{2nd}}
 = \frac{g^2}{4\pi}
    \sqrt{2}M_\mathrm{s}
    \frac{N}{60} \,,
\end{equation}
 where $N=3$ (see Appendix \ref{sec:amplitude_and_width}).

We include interference effects
 between first and second string excited states.
The squared amplitudes in eqs.(\ref{eq:1st_excited_squared_amplitude_qg_t-channel})-(\ref{eq:1st_excited_squared_amplitude_qg_u-channel})
 and eqs.(\ref{eq:2nd_excited_squared_amplitude_t-channel})-(\ref{eq:2nd_excited_squared_amplitude_u-channel})
 are formulae in a good approximation only around
 $\hat{s} \simeq M_\mathrm{s}^2$
 and
 $\hat{s} \simeq 2M_\mathrm{s}^2$,
 respectively.
The interference effects should be calculated
 without replacements of $\hat{s}$ by $M_\mathrm{s}^2$
 or $2M_\mathrm{s}^2$
 in the processes of deriving
 eqs.(\ref{eq:1st_excited_squared_amplitude_qg_t-channel})-(\ref{eq:1st_excited_squared_amplitude_qg_u-channel})
 and eqs.(\ref{eq:2nd_excited_squared_amplitude_t-channel})-(\ref{eq:2nd_excited_squared_amplitude_u-channel}), respectively
 (see Appendix \ref{sec:amplitude_and_width} for details).

The structure of the amplitudes for the processes of $qg \rightarrow qg$
 is obtained as
\begin{equation}
 A_{qg \rightarrow qg}
 = A^{J=1/2}_{\rm 1st} + A^{J=1/2}_{\rm 2nd} + A^{J=3/2}_{\rm 2nd} \,,
\end{equation}
 where $A^J_{n{\rm th}}$ describe contributions
 of $n$th string excited states with spin $J$.
The squared amplitude includes three kinds of interference terms.
\begin{equation}
	\begin{split}
 \bigl\vert A_{qg \rightarrow qg} \bigl\vert^2
 = & \bigl\vert A^{J=1/2}_{\rm 1st} \bigl\vert^2
 + \bigl\vert A^{J=1/2}_{\rm 2nd} \bigl\vert^2
 + \bigl\vert A^{J=3/2}_{\rm 2nd} \bigl\vert^2
 \\
 + & \biggl( A^{J=1/2}_{\rm 1st} {A^{J=1/2}_{\rm 2nd}}^* + \rm{c.c} \biggr)
 + \biggl( A^{J=1/2}_{\rm 1st} {A^{J=3/2}_{\rm 2nd}}^* + \rm{c.c} \biggr)
 + \biggl( A^{J=1/2}_{\rm 2nd} {A^{J=3/2}_{\rm 2nd}}^* + \rm{c.c} \biggr) \,.
	\end{split}
\end{equation}
The last term, an interference between two amplitudes
 with poles at the same place, $M_{\rm{2nd}}^2 = 2M_{\rm s}^2$,
 gives large contributions to the total peak cross section
 at $\hat{s} = 2M_{\rm s}^2$.
The other two interference terms give small suppression effects
 in the region of $M_{\rm s}^2 < \hat{s} < 2M_{\rm s}^2$.
The interference effects between the SM contributions and string contributions
 are not included.
There should be some effects
 due to $\hat{u}$- and $\hat{t}$-channel exchanges of string excited states,
 which might give a certain effect
 in the region of $M_{\rm s}^2 < \hat{s} < 2M_{\rm s}^2$.
We leave detailed analyses in these effects for future works.
Without including these effects,
 the peak cross sections
 at $\hat{s} = M_{\rm s}$ and $\hat{s} = 2M_{\rm s}^2$
 should be correctly estimated,
 which is sufficient for the aim of this paper.

\begin{figure}[t]
	\centering
	\includegraphics[width=12cm]
					  {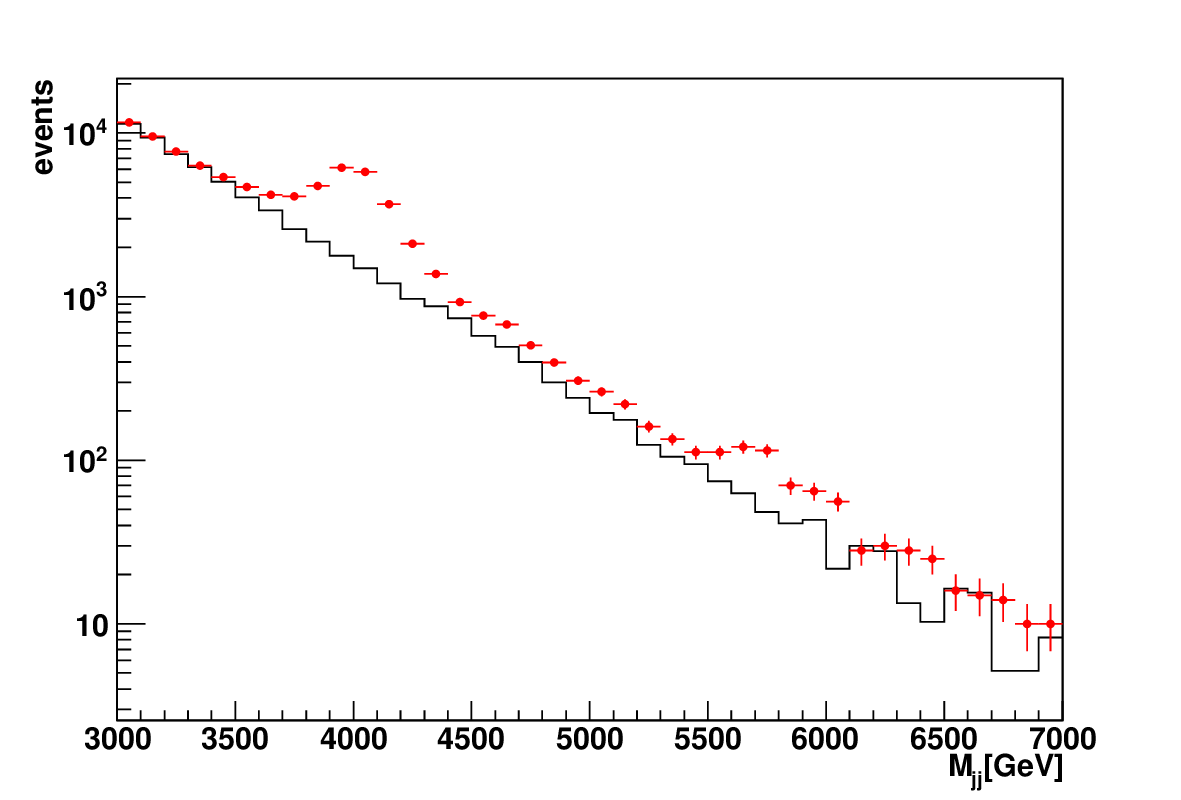}
	\caption{The dijet invariant mass distribution
			   with interference effects between first and second excited states
			   for $M_{\rm s} = 4\,{\rm TeV}$
			   with $50\,\mathrm{fb}^{-1}$ of integrated luminosity 
			   at $\sqrt{s} = 14\,{\rm TeV}$. 
			  Black histogram shows the SM background.}
	\label{fig:1st_and_2nd_excited_dijet_mass_50fb-1}
\end{figure}

Event samples
 for the SM with first and second string excited states are generated.
The number of events is corresponding to $50\,{\rm fb}^{-1}$
 at $\sqrt{s}=14\,{\rm TeV}$.
Kinematical cuts imposed in this analysis are
\begin{equation}
	\label{eq:analysis_cut_2}
	\begin{split}
 & \hspace{11mm}
 p_{{\rm T},j_1}
 > 350 \,{\rm GeV} \,, \\
 & |y_{1,2}|
     < 2.3 \,, \hspace{6mm}
    |y_1-y_2|
     < 1.7 \,.
	\end{split}
\end{equation}
The dijet invariant mass distribution
 is shown in Fig.\ref{fig:1st_and_2nd_excited_dijet_mass_50fb-1}.
We see the first string resonance at
 $M_{jj} = M_{\rm s} = 4\,{\rm TeV}$
 and the second string resonance at
 $M_{jj} = \sqrt{2} \times M_{\rm s} \simeq 5.66\,{\rm TeV}$.
We need integrated luminosity of $50\,\mathrm{fb}^{-1}$
 to obtain enough events at high energies
 and to see the second string resonance clearly.
Signal-to-noise ratios for the second string resonance
 are calculated in the dijet invariant mass window
 $[\sqrt{2}M_\mathrm{s}-250\,\mathrm{GeV},\sqrt{2}M_\mathrm{s}+250\,\mathrm{GeV}]$. 
For $18.5\,\mathrm{fb}^{-1}$ of integrated luminosity,
 $S/\sqrt{B}=71/\sqrt{133}\simeq6\sigma$,
 and for $50\,\mathrm{fb}^{-1}$,
 $S/\sqrt{B}=217/\sqrt{313}\simeq12\sigma$.

If a resonance in the dijet invariant mass distribution
 is observed at the LHC,
 we should look for a second resonance
 at rather close to the first resonance.
If the second resonance is discovered
 at a place $\sqrt{2}$ times far from the place of the first resonance,
 which is a strong signature for low-scale string models.

\section{Conclusions}
	\label{sec:conclusion}

	\hspace{5mm}
We have investigated phenomenology of low-scale string models at the LHC.
Since in low-scale string models,
 the string scale $M_s$ is of the order of TeV with large extra dimensions,
 it is expected some string excited states
 as well as KK states
 may be observed at the LHC.
It has been known that
 parton two-body scattering processes at the LHC
 is highly model independent,
 namely independent from the way of compactification of extra dimensions,
 and processes with single KK state in $s$-channel
 are highly suppressed.
Therefore, dijet events at the LHC
 are that should be focused to observe string excited states.

The string excited states appear as resonances
 in the dijet invariant mass distribution.
Since several string excited states with different spins
 are degenerated in mass of $M_s$,
 dijet events on the resonance exhibit special angular distributions
 which are not realized in any other ``new physics'' with a single heavy state.
We have investigated the $\chi$-distribution on the resonance
 by Monte Carlo simulations
 and shown that main contributions
 of two kinds of string excited states of quarks with spin $J=1/2$ and $J=3/2$
 can be confirmed with 20 fb$^{-1}$ of integrated luminosity
 at $\sqrt{s} = 14\,{\rm TeV}$.

Another inevitable prediction of low-scale string models
 is emergence of a second resonance
 in the dijet invariant mass distribution
 due to second string excited states of quarks and gluons.
A distinct property is the place of the second resonance,
 namely the mass of second string excited states.
The masses should be $\sqrt{2}M_s$
 which should be compared with
 that masses of second KK states is twice that of first KK states.
We have shown by Monte Carlo simulations
 that the second resonance in the dijet invariant mass distribution
 due to second string excited states
 can be observed with 50 fb$^{-1}$ of integrated luminosity
 at $\sqrt{s} = 14\,{\rm TeV}$.

Though these are necessary and inevitable signatures
 of low-scale string models,
 more information is necessary to establish low-scale string.
For example, confirmation of special angular distributions
 of dijet events on the second resonance,
 and emergence of light anomalous U$(1)$ gauge bosons
 with special interactions related with anomaly structures
 (though this is rather highly model dependent)
 should be worth to investigate in future.
We are planning a systematic and detailed study of second string excited states
 with larger values of string scales than $4\,\mathrm{TeV}$
 which is the current lower bound by the CMS experiment.

\vspace{3mm}
\textbf{Note added.} 
After accepting this article for publication, we became aware
 that the importance of second excited states has been pointed out in ref. \cite{Dong:2010jt}.

\section*{Acknowledgements}
	\label{sec:acknowledgement}

	\hspace{5mm}
The authors would like to thank Koji Terashi
 for valuable advices on the experiment of dijet physics
 and techniques of Monte Carlo simulations.
The authors would like to thank Junpei Maeda
 for advices on the techniques of Monte Carlo simulations.
N.K. is supported in part by Grand-in-Aid for Research No.\,DC203
 from Tokyo Metropolitan University.

\appendix

\section{Amplitudes and widths with string excited states}
	\label{sec:amplitude_and_width}

	\hspace{5mm}
Open-string two-body scattering amplitudes between quarks and gluons
 are calculated by the world-sheet superconformal field theory
 in flat space-time as
\begin{equation}
	\label{eq:string_amplitude_qg_spin1-2}
\mathcal{M} ( q_{1}^\pm g_{2}^\pm \rightarrow q_{3}^\pm g_{4}^\pm )
= 2g^2
   \sqrt{-\frac{s}{u}}
    	\biggl[
    		\frac{s}{t}
		V_s
		\bigl(T^{a_2}T^{a_4}\bigr)_{\alpha_3\alpha_1}
	    + \frac{u}{t}
		V_u
		\bigl(T^{a_4}T^{a_2}\bigr)_{\alpha_3\alpha_1}
	\biggr] \,,
\end{equation}
\begin{equation}
	\label{eq:string_amplitude_qg_spin3-2}
 \mathcal{M} ( q_{1}^\pm g_{2}^\mp \rightarrow q_{3}^\pm g_{4}^\mp )
 = 2g^2
    \sqrt{-\frac{u}{s}}
	\biggl[
		\frac{s}{t}
		V_s
		\bigl(T^{a_2}T^{a_4}\bigr)_{\alpha_3\alpha_1}
	    + \frac{u}{t}
	       V_u
	       \bigl(T^{a_4}T^{a_2}\bigr)_{\alpha_3\alpha_1}
	\biggr] \,,
\end{equation}
 where $\alpha_1$, $\alpha_3$ are color indices of quarks,
 $a_2$, $a_4$ are color indices of gluons,
 and $s$, $t$ and $u$ are the Mandelstam variables of partons
 with $s+t+u=0$.
Here, the generators $T^{a_i}$ are
 for the fundamental representation of SU$(3)_\mathrm{color}$ gauge group.
Amplitudes
 $\mathcal{M} ( q_{1}^\pm g_{2}^\pm \rightarrow g_{3}^\pm q_{4}^\pm )$
 and $\mathcal{M} ( q_1^\pm g_2^\mp \rightarrow g_3^\mp q_4^\pm )$
 are obtained by replacements of $u \leftrightarrow t$ and $3 \leftrightarrow4$
 in the above amplitudes
 of eqs.(\ref{eq:string_amplitude_qg_spin1-2})-(\ref{eq:string_amplitude_qg_spin3-2}).

The functions $V_{s,t,u}$
 in eqs.(\ref{eq:string_amplitude_qg_spin1-2})-(\ref{eq:string_amplitude_qg_spin3-2})
 are ``form factor" functions of the Mandelstam variables
 which represent string effects in open-string amplitudes.
\begin{equation}
	\label{eq:V-function_all}
 V_t = V(s,t,u) \,, \hspace{6mm}
 V_u = V(t,u,s) \,, \hspace{6mm}
 V_s = V(u,s,t) \,,
\end{equation}
where
\begin{equation}
	\label{eq:V-function_2}
 V(s,t,u)
 = \frac{\Gamma(1-s/M_\mathrm{s}^2) \Gamma(1-u/M_\mathrm{s}^2)}
             {\Gamma(1+t/M_\mathrm{s}^2)}
 = \frac{1}{M_\mathrm{s}^2}
    \frac{su}{t}
    \frac{\Gamma(-s/M_\mathrm{s}^2) \Gamma(-u/M_\mathrm{s}^2)}
             {\Gamma(t/M_\mathrm{s}^2)} \,.
\end{equation}
These functions are expanded by sums over infinite $s$-channel poles.
\begin{equation}
	\label{eq:V-function_expanded_2}
 V_t
 = \frac{1}{M_\mathrm{s}^2}
    \frac{su}{t}
 	\biggl\{
    	     - \sum^{\infty}_{n=1}
			\frac{1}{n!}
			\frac{1}{(M_\mathrm{s}^2)^{n-1}}
			\frac{1}{s - nM_\mathrm{s}^2}
		\prod^n_{J=1}
			\bigl( u + JM_\mathrm{s}^2 \bigr)
     + \mathcal{O}\bigl( (s - nM_\mathrm{s}^2)^ 0\bigr)
     \biggr\} \,.
\end{equation}
Note that the function $V_s$ has no $s$-channel poles.
The terms of
 $\mathcal{O} \bigl( (s - nM_\mathrm{s}^2)^0 \bigr)$
 in eq.(\ref{eq:V-function_expanded_2})
 can be neglected if we consider only near each $n$th pole:
 $s \simeq nM_\mathrm{s}^2$.
The each pole corresponds to $n$th string excited state
 with mass $M_n = \sqrt{n} M_{\rm s}$.

The form factor functions are also expanded by $(M_\mathrm{s}^2)^{-1}$ as
\begin{equation}
	\label{eq:low_energy_limit}
 V_t
 = 1 - \frac{\pi^2}{6} su \biggl( \frac{1}{M_\mathrm{s}^2} \biggr)^2
       + \mathcal{O} \bigl( (M_\mathrm{s}^2)^{-3} \bigr) \,,
\end{equation}
 which corresponds to the low-energy limit,
 namely $V_{s,t,u} \rightarrow 1$ with $M_\mathrm{s} \rightarrow \infty$.
In a low-energy limit, the open-string amplitudes of $qg \rightarrow qg$
 in eqs.(\ref{eq:string_amplitude_qg_spin1-2})-(\ref{eq:string_amplitude_qg_spin3-2})
 exactly coincide with the QCD amplitudes,
\begin{equation}
	\label{eq:QCD_amplitude_qg_spin1-2}
 \mathcal{M}_\mathrm{QCD} ( q_{1}^\pm g_{2}^\pm \rightarrow q_{3}^\pm g_{4}^\pm)
 = 2g^2
    \sqrt{-\frac{s}{u}}
 	\biggl[
		\frac{s}{t}
		\bigl(T^{a_2}T^{a_4}\bigr)_{\alpha_3\alpha_1}
	    + \frac{u}{t}
		\bigl(T^{a_4}T^{a_2}\bigr)_{\alpha_3\alpha_1}
	\biggr] \,,
\end{equation}
\begin{equation}
	\label{eq:QCD_amplitude_qg_spin3-2}
 \mathcal{M}_\mathrm{QCD} ( q_{1}^\pm g_{2}^\mp \rightarrow q_{3}^\pm g_{4}^\mp)
 = 2g^2
    \sqrt{-\frac{u}{s}}
	\biggl[
		\frac{s}{t}
		\bigl(T^{a_2}T^{a_4}\bigr)_{\alpha_3\alpha_1}
	    + \frac{u}{t}
	    	\bigl(T^{a_4}T^{a_2}\bigr)_{\alpha_3\alpha_1}
	\biggr] \,.
\end{equation}

\subsection{The first excited states}
	\label{sec:qg_1st}

	\hspace{5mm}
In case of $n=1$, the form factor functions are approximated
 near the first $s$-channel pole, $s \simeq M_\mathrm{s}^2$, as
\begin{equation}
	\label{eq:V-function_near_1st_pole}
 V_t \simeq \frac{u}{s - M_\mathrm{s}^2} \,, \hspace{6mm}
 V_u \simeq \frac{t}{s - M_\mathrm{s}^2} \,.
\end{equation}
Amplitudes of $qg \rightarrow qg$ with exchanges of the first quark excited states
 in eqs.(\ref{eq:string_amplitude_qg_spin1-2})-(\ref{eq:string_amplitude_qg_spin3-2})
 are recast into
\begin{equation}
	\label{eq:1st_amplitude_qg_spin1-2}
 \mathcal{M}_\mathrm{1st} ( q_1^\pm g_2^\pm \rightarrow q_3^\pm g_4^\pm )
 \simeq -2g^2 M_\mathrm{s}^2
 		  \frac{1}{s-M_\mathrm{s}^2}
		  d^{J=1/2}_{\mp1/2,\mp1/2}(\theta)
		  \bigl(T^{a_4}T^{a_2}\bigr)_{\alpha_3\alpha_1} \,,
\end{equation}
\begin{equation}
	\label{eq:1st_amplitude_qg_spin3-2}
 \mathcal{M}_\mathrm{1st} ( q_1^\pm g_2^\mp \rightarrow q_3^\pm g_4^\mp )
 \simeq -2g^2 M_\mathrm{s}^2
 		 \frac{1}{s-M_\mathrm{s}^2}
		 d^{J=3/2}_{\pm3/2,\pm3/2}(\theta)
		 \bigl(T^{a_4}T^{a_2}\bigr)_{\alpha_3\alpha_1} \,,
\end{equation}
 where $\mathrm{d}^J_{J_z,J_z^\prime}(\theta)$s are the Wigner $d$-functions,
\begin{equation}
	\label{eq:d-function_spin1-2_and_spin3-2}
 d^{J=1/2}_{\pm1/2,\pm1/2}(\theta)
 = \cos\frac{\theta}{2} \,, \hspace{6mm}
 d^{J=3/2}_{\pm3/2,\pm3/2}(\theta)
 = \cos\frac{\theta}{2} \biggl(\frac{1+\cos\theta}{2}\biggr) \,.
\end{equation}
The Wigner $d$-function $d^J_{J_z,J_z^\prime}(\theta)$ represents
 angular dependence of a state with spin $J$
 with conditions of the initial spin component $J_z$ along the $z$-axis
 and the final spin component $J_z^\prime$ along the $z^\prime$-axis.
The angle $\theta$ is that between the $z$-axis and $z^\prime$-axis.
Therefore, the exchanged state
 in the process of eq.(\ref{eq:1st_amplitude_qg_spin1-2}) with $J_z = \pm1/2$
 are first quark excited states with $J = 1/2$
 and the exchanged states
 in the process of eq.(\ref{eq:1st_amplitude_qg_spin3-2}) with $J_z = \pm3/2$
 are first quark excited state with $J = 3/2$.

The $s$-channel poles in eq.(\ref{eq:V-function_expanded_2})
 have to be softened into Breit-Wigner forms:
\begin{equation}
	\label{eq:Breit-Wigner_form}
 \frac{1}{s-nM_\mathrm{s}^2}
 \rightarrow \frac{1}{ s - nM_\mathrm{s}^2
                                   + i \sqrt{n}M_\mathrm{s} \Gamma^J_{q^*,n\mathrm{th}} } \,,
\end{equation}
 where the widths of the $n$th excited states of quarks $q^*$ with spin $J$,
 $\Gamma^J_{q^*,n\mathrm{th}}$,
 are added by hand.

The width is calculated as follows.
A Lorentz invariant amplitude in a two-body scattering process,
 in which a state with mass $M$, spin $J$ and the gauge index $\alpha$
 is exchanged in $s$-channel,
 is written in the following form,
\begin{equation}
	\label{eq:invariant_amplitude}
 \mathcal{M}
 = -\sum_{\alpha,J}
      \frac{1}{s-M^2}
      F^{\alpha J}_{\lambda_3,\lambda_4;\alpha_3,a_4}
      F^{\alpha J}_{\lambda_1,\lambda_2;\alpha_1,a_2}
      d^J_{\lambda_1-\lambda_2,\lambda_3-\lambda_4}(\theta) \,,
\end{equation}
 where the entire negative sign is just a convention.
The quantities $F^{\alpha J}_{\lambda_i,\lambda_j;\alpha_i,a_j}$
 are vertex factors
 in decay of the exchanged state
 in the process of eq.(\ref{eq:invariant_amplitude})
 with spin $J$ and the gauge index $\alpha$
 into two states with the helicities $\lambda_i$, $\lambda_j$
 and gauge indices $\alpha_i$, $a_j$,
 which are independent from
 a spin component $J_z$ of the exchanged state.
The decay width of the exchanged state can be calculated as
\begin{equation}
	\label{eq:decay_width_2}
 \Gamma^{\alpha J}_{\lambda_3,\lambda_4;\alpha_3,a_4} 
 = \frac{1}{16\pi M}
    \frac{1}{2J+1}
    \bigl| F^{\alpha J}_{\lambda_3,\lambda_4;\alpha_3,a_4} \bigr|^2 \,.
\end{equation}
See Ref.\cite{Anchordoqui:2008hi} for details.

From the amplitudes
 of eqs.(\ref{eq:1st_amplitude_qg_spin1-2})-(\ref{eq:1st_amplitude_qg_spin3-2}),
 we have
\begin{equation}
	\label{eq:1st_vertex_coefficient_spin1-2_and_spin3-2}
 F^{\alpha J=1/2}_{\pm1/2,\pm1;\alpha_3,a_4}\biggr|_\mathrm{1st}
 = \sqrt{2}
    gM_\mathrm{s}
    \bigl(T^{a_4}\bigr)_{\alpha\alpha_3} \,, \hspace{6mm}
 F^{\alpha J=3/2}_{\pm1/2,\mp1;\alpha_3,a_4}\biggr|_\mathrm{1st}
 = \sqrt{2}
    gM_\mathrm{s}
    \bigl(T^{a_4}\bigr)_{\alpha\alpha_3} \,.
\end{equation}
Color-averaged decay widths of the first quark excited state $q^*$ with $J$
 in a decay process of $q^*\rightarrow qG$ are calculated as
\begin{equation}
	\label{eq:1st_width_q}
 \Gamma^{J}_{q^*\rightarrow qG,\mathrm{1st}}
 = \frac{1}{16\pi M_\mathrm{s}}
    \frac{1}{2J+1}
    \frac{1}{N}
    \sum_\alpha
    \sum_{\alpha_3,a_4}
    \sum_{\lambda_3,\lambda_4}
    	\bigl| F^{\alpha J}_{\lambda_3,\lambda_4;\alpha_3,a_4} \bigr|^2,
\end{equation}
where $N=3$.
In case of $J=1/2$,
\begin{equation}
	\begin{split}
	\label{eq:1st_width_q_spin1-2}
 \Gamma^{J=1/2}_{q^*\rightarrow qG,\mathrm{1st}} 
 & = \frac{1}{16\pi M_\mathrm{s}}
 	 \frac{1}{2\times\frac{1}{2}+1}
	 \frac{1}{N}
	 \sum_\alpha
	 \sum_{\alpha_3,a_4}
	 	\frac{1}{2}
		\biggl\{
			\bigl| F^{\alpha J=1/2}_{+1/2,+1;\alpha_3,a_4} \bigr|^2
		    + \bigl| F^{\alpha J=1/2}_{-1/2,-1;\alpha_3,a_4} \bigr|^2
		\biggr\} \\
 & = \frac{g^2M_\mathrm{s}}{16\pi}
	 \frac{1}{N}
	 \sum_{a_4}
	 	\mathrm{tr}\bigl(T^{a_4}T^{a_4}\bigr) \,.
	\end{split}
\end{equation}
The factor $\frac{1}{2}$ of a sum over helicities comes from a fact
 that the final state helicity configuration 
 $(\lambda_3,\lambda_4)=(+1/2,+1)$
 couples only to the initial state helicity $J_z=-1/2$,
 while 
 $(\lambda_3,\lambda_4)=(-1/2,-1)$
 couples only to $J_z=+1/2$.
The same applies to the case of $J=3/2$
 (see eqs.(\ref{eq:1st_amplitude_qg_spin1-2})-(\ref{eq:1st_amplitude_qg_spin3-2})).

The decay products $G$ in the process of $q^* \rightarrow qG$
 are gauge bosons of U$(3)_\mathrm{color}$ gauge group.
The U$(N)$ gauge bosons $G^A$ $(A=1,\cdots,N^2)$
 are split into SU$(N)$ gauge bosons $g^a$ $(a=1,\cdots,N^2-1)$,
 namely gluons in case of $N=3$
 and a U$(1)$ gauge boson $C^0$.
The U$(N)$ generators $T^A$
 are also split into SU$(N)$ generators $T^a$
 with $\mathrm{tr}(T^aT^b) = \frac{1}{2}\delta^{ab}$
 and a U$(1)$ generator $T^0 = \frac{1}{\sqrt{2N}} \textbf{1}_N$.
Hence,
\begin{equation}
	\begin{split}
	\label{eq:trace_generator}
 \frac{1}{N}
 \sum_{a=1}^{N^2-1}
 	\mathrm{tr}\bigl(T^aT^a\bigr)
 = \frac{N^2-1}{2N}
	\hspace{6mm} 
	 & {\rm for} \, q^* \rightarrow qg \,, \\
 \frac{1}{N}
 \mathrm{tr}\bigl(T^0T^0\bigr)
 = \frac{1}{2N}
 	\hspace{12.5mm}
	 & {\rm for} \, q^*\rightarrow qC^0 \,.
	\end{split}
\end{equation}
Total widths of the first quark excited states with $J=1/2$ and $J=3/2$
 are obtained as eq.(\ref{eq:1st_excited_width_q}).

Spin- and color-averaged squared amplitude of $qg \rightarrow qg$
 with exchanges of the first quark excited states with $J=1/2$ and $J=3/2$
 is calculated
 from eqs.(\ref{eq:1st_amplitude_qg_spin1-2})-(\ref{eq:1st_amplitude_qg_spin3-2}) as
\begin{equation}
	\begin{split}
	\label{eq:1st_squared_amplitude_qg}
 \bigl| \mathcal{M}_\mathrm{1st} ( q_1 g_2 \rightarrow q_3 g_4 ) \bigr|^2
 & = \frac{1}{N}
 	 \frac{1}{N^2-1}
	 \sum_{\alpha_1,a_2}
	 \sum_{\alpha_3,a_4} \\
 & \hspace{5mm}
 	 	\times
		\biggl(\frac{1}{2}\biggr)^2
		\biggl\{
			\bigl| \mathcal{M}_\mathrm{1st}
				    ( q_1^+ g_2^+ \rightarrow q_3^+ g_4^+ ) \bigr|^2
		    + \bigl| \mathcal{M}_\mathrm{1st}
		    		    ( q_1^- g_2^- \rightarrow q_3^- g_4^- ) \bigr|^2 \\
 & \hspace{19mm}
 		    + \bigl| \mathcal{M}_\mathrm{1st}
		    		    ( q_1^+ g_2^- \rightarrow q_3^+ g_4^- ) \bigr|^2
		    + \bigl| \mathcal{M}_\mathrm{1st}
		    		    ( q_1^- g_2^+ \rightarrow q_3^- g_4^+ ) \bigr|^2
		\biggr\} \,.
	\end{split}
\end{equation}
This is equal to the right-handed side of the second line
 of eq.(\ref{eq:1st_excited_squared_amplitude_qg_t-channel}).

\subsection{The second excited states}
	\label{sec:qg_2nd}

	\hspace{5mm}
In case of $n=2$,
 the form factor functions are approximated near the second $s$-channel pole,
 $s \simeq 2M_\mathrm{s}^2$, as
\begin{equation}
	\label{eq:V-function_near_2nd_pole}
 V_t
 \simeq \frac{1}{M_\mathrm{s}^2}
 		 \frac{u}{s - 2M_\mathrm{s}^2}
		 \biggl( u + \frac{s}{2} \biggr) \,, \hspace{6mm}
 V_t
 \simeq \frac{1}{M_\mathrm{s}^2}
 		 \frac{t}{s - 2M_\mathrm{s}^2}
		 \biggl( t + \frac{s}{2} \biggr) \,.
\end{equation}
Amplitudes of $qg \rightarrow qg$
 with exchanges of the second quark excited states are recast into
\begin{equation}
	\label{eq:2nd_amplitude_qg_spin1-2}
 \mathcal{M}_\mathrm{2nd} ( q_1^\pm g_2^\pm \rightarrow q_3^\pm g_4^\pm )
 \simeq -4g^2 M_\mathrm{s}^2
 		 \frac{1}{s-2M_\mathrm{s}^2}
		 \biggl[
		 	\frac{1}{3}
			d^{J=1/2}_{\mp1/2,\mp1/2}(\theta)
		    + \frac{2}{3}
		    	d^{J=3/2}_{\mp1/2,\mp1/2}(\theta)
		 \biggr]
		 \bigl(T^{a_4}T^{a_2}\bigr)_{\alpha_3\alpha_1} \,,
\end{equation}
\begin{equation}
	\label{eq:2nd_amplitude_qg_spin3-2}
 \mathcal{M}_\mathrm{2nd} ( q_1^\pm g_2^\mp \rightarrow q_3^\pm g_4^\mp )
 \simeq -4g^2 M_\mathrm{s}^2
 		 \frac{1}{s-2M_\mathrm{s}^2}
		 \biggl[
		 	\frac{3}{5}
			d^{J=3/2}_{\pm3/2,\pm3/2}(\theta)
		    + \frac{2}{5}
		  	d^{J=5/2}_{\pm3/2,\pm3/2}(\theta)
		 \biggr]
		 \bigl(T^{a_4}T^{a_2}\bigr)_{\alpha_3\alpha_1} \,,
\end{equation}
 where the Wigner $d$-functions
 are given in eq.(\ref{eq:d-function_spin1-2_and_spin3-2}) and
\begin{equation}
	\label{eq:d-function_spin3-2_and_spin5-2}
 d^{J=3/2}_{\pm1/2,\pm1/2}(\theta)
 = \cos\frac{\theta}{2}
    \biggl( \frac{3\cos\theta-1}{2} \biggr) \,, \hspace{6mm}
 d^{J=5/2}_{\pm3/2,\pm3/2}(\theta)
 = \cos\frac{\theta}{2}
    \biggl( \frac{1+\cos\theta}{2} \biggr)
    \biggl( \frac{5\cos\theta-3}{2} \biggr) \,.
\end{equation}
Therefore, the exchanged states
 in the process of eq.(\ref{eq:2nd_amplitude_qg_spin1-2}) with $J_z=\pm1/2$
 are second quark excited states with $J=1/2$ and $J=3/2$,
 and the exchanged states
 in the process of eq.(\ref{eq:2nd_amplitude_qg_spin3-2}) with $J_z=\pm3/2$
 are second quark excited states with $J=3/2$ and $J=5/2$.

From the amplitudes
 of eqs.(\ref{eq:2nd_amplitude_qg_spin1-2})-(\ref{eq:2nd_amplitude_qg_spin3-2}),
 we have
\begin{equation}
	\label{eq:2nd_vertex_coefficient_spin1-2_and_spin3-2}
 F^{\alpha J=1/2}_{\pm1/2,\pm1;\alpha_3,a_4}\biggr|_\mathrm{2nd}
 = \frac{2}{\sqrt{3}}
    gM_\mathrm{s}
    \bigl(T^{a_4}\bigr)_{\alpha\alpha_3} \,, \hspace{6mm}
 F^{\alpha J=3/2}_{\pm1/2,\pm1;\alpha_3,a_4}\biggr|_\mathrm{2nd}
 = 2 \sqrt{\frac{2}{3}}
    gM_\mathrm{s}
    \bigl(T^{a_4}\bigr)_{\alpha\alpha_3} \,,
\end{equation}
\begin{equation}
	\label{eq:2nd_vertex_coefficient_spin3-2_and_spin5-2}
 F^{\alpha J=3/2}_{\pm1/2,\mp1;\alpha_3,a_4}\biggr|_\mathrm{2nd}
 = 2 \sqrt{\frac{3}{5}}
    gM_\mathrm{s}
    \bigl(T^{a_4}\bigr)_{\alpha\alpha_3} \,, \hspace{6mm}
 F^{\alpha J=5/2}_{\pm1/2,\mp1;\alpha_3,a_4}\biggr|_\mathrm{2nd}
 = 2 \sqrt{\frac{2}{5}}
    gM_\mathrm{s}
    \bigl(T^{a_4}\bigr)_{\alpha\alpha_3} \,. \hspace{2mm}
\end{equation}
Color-averaged decay widths of the second quark excited state $q^*$ with $J$
 are calculated.
In case of $J=3/2$,
 since there are contributions from both processes
 of eq.(\ref{eq:2nd_amplitude_qg_spin1-2}) and eq.(\ref{eq:2nd_amplitude_qg_spin3-2})
 with exchanges of states with $J_z=\pm1/2$ and $J_z=\pm3/2$, respectively,
\begin{equation}
	\begin{split}
	\label{eq:2nd_width_q_spin3-2}
 \Gamma^{J=3/2}_{q^*,\mathrm{2nd}}
 & = \frac{1}{16\pi\sqrt{2}M_\mathrm{s}}
 	 \frac{1}{2\times\frac{3}{2}+1}
	 \frac{1}{N}
	 \sum_\alpha
	 \sum_{\alpha_3,a_4}
	 	\frac{1}{2}
		\biggl\{
			\bigl| F^{\alpha J=3/2}_{+1/2,+1;\alpha_3,a_4} \bigr|^2
		    + \bigl| F^{\alpha J=3/2}_{-1/2,-1;\alpha_3,a_4} \bigr|^2 \\
 & \hspace{63mm} 
 		    + \bigl| F^{\alpha J=3/2}_{+1/2,-1;\alpha_3,a_4} \bigr|^2
		    + \bigl| F^{\alpha J=3/2}_{-1/2,+1;\alpha_3,a_4} \bigr|^2
		 \biggr\} \\
 & = \frac{g^2M_\mathrm{s}}{16\sqrt{2}\pi}
 	 \biggl( \frac{2}{3} + \frac{3}{5} \biggr)
	 \frac{1}{N}
	 \sum_{a_4}
	 	\mathrm{tr}\bigl(T^{a_4}T^{a_4}\bigr) \,.
	\end{split}
\end{equation}
Total widths of the second quark excited states with $J=1/2$, $J=3/2$ and $J=5/2$
 are obtained as eq.(\ref{eq:2nd_excited_width}).
We are neglecting subdominant decay processes to the first string excited states.

The spin- and color-averaged squared amplitude is calculated
 from eqs.(\ref{eq:2nd_amplitude_qg_spin1-2})-(\ref{eq:2nd_amplitude_qg_spin3-2})
 in a similar way to eq.(\ref{eq:1st_squared_amplitude_qg})
 and the result is equal to eq.(\ref{eq:2nd_excited_squared_amplitude_t-channel}).

\subsection{The interference between the first and second excited states}
	\label{sec:qg_interference}

	\hspace{5mm}
The interference effects between first and second excited states
 have to be calculated
 without replacements of $s$ with $M_\mathrm{s}^2$
 in eqs.(\ref{eq:1st_amplitude_qg_spin1-2})-(\ref{eq:1st_amplitude_qg_spin3-2})
 and with $2M_\mathrm{s}^2$
 in eqs.(\ref{eq:2nd_amplitude_qg_spin1-2})-(\ref{eq:2nd_amplitude_qg_spin3-2}).
It is because we need amplitudes in the region of $s$
 a little far from the poles of first and second excited states.

Amplitudes of $qg \rightarrow qg$
 with exchanges of the first and second quark excited sates are written as
\begin{equation}
	\begin{split}
	\label{eq:1st_and_2nd_amplitude_qg_spin1-2}
 & \mathcal{M}_\mathrm{1st+2nd} ( q_1^\pm g_2^\pm \rightarrow q_3^\pm g_4^\pm )
 = 2g^2
    \bigl(T^{a_4}T^{a_2}\bigr)_{\alpha_3\alpha_1} \\
 & \times
     \Biggl\{
             - \frac{s^2}{M_\mathrm{s}^2}
		 \sqrt{-\frac{u}{s}}
		 \frac{1}{ s - M_\mathrm{s}^2
		 		+ i M_\mathrm{s} \Gamma^{J=1/2}_{q^*,\mathrm{1st}}} \\
 & \hspace{7mm}
 	      - \frac{s^3}{4M_\mathrm{s}^4}
		 \Biggl[
			\frac{1}{3}
			\sqrt{-\frac{u}{s}}
			\frac{1}{ s - 2M_\mathrm{s}^2
				      + i \sqrt{2}M_\mathrm{s} \Gamma^{J=1/2}_{q^*,\mathrm{2nd}}} 
		    + \frac{2}{3}
		    	\sqrt{-\frac{u}{s}}
			\biggl(\frac{3t+s}{s}\biggr)
			\frac{1}{ s - 2M_\mathrm{s}^2
				      + i \sqrt{2}M_\mathrm{s} \Gamma^{J=3/2}_{q^*,\mathrm{2nd}}}
		  \Biggr]
    \Biggr\} \,,
	\end{split}
\end{equation}
\begin{equation}
	\label{eq:1st_and_2nd_amplitude_qg_spin3-2}
	\begin{split}
 & \mathcal{M}_\mathrm{1st+2nd} ( q_1^\pm g_2^\mp \rightarrow q_3^\pm g_4^\mp)
 = 2g^2
    \bigl(T^{a_4}T^{a_2}\bigr)_{\alpha_3\alpha_1} \\
 & \times
     \Biggl\{
     	     - \frac{s^2}{M_\mathrm{s}^2}
		\biggl(-\frac{u}{s}\biggr)^{3/2}
		\frac{1}{ s - M_\mathrm{s}^2
			      + i M_\mathrm{s} \Gamma^{J=3/2}_{q^*,\mathrm{1st}}} \\
 & \hspace{7mm}
 	     - \frac{s^3}{4M_\mathrm{s}^4}
		\Biggl[
			\frac{3}{5}
			\biggl(-\frac{u}{s}\biggr)^{3/2}
			\frac{1}{ s - 2M_\mathrm{s}^2
				      + i \sqrt{2}M_\mathrm{s} \Gamma^{J=3/2}_{q^*,\mathrm{2nd}}} 
		    + \frac{2}{5}
		       \biggl(-\frac{u}{s}\biggr)^{3/2}
		       \biggl(\frac{5t+s}{s}\biggr)
		       \frac{1}{ s - 2M_\mathrm{s}^2
		       		      + i \sqrt{2}M_\mathrm{s} \Gamma^{J=5/2}_{q^*,\mathrm{2nd}}}
		\Biggr]
     \Biggr\} \,.
	\end{split}
\end{equation}
Spin- and color-averaged squared amplitude
 including the interference effects is calculated from
 eqs.(\ref{eq:1st_and_2nd_amplitude_qg_spin1-2})-(\ref{eq:1st_and_2nd_amplitude_qg_spin3-2})
 as
\begin{equation}
	\begin{split}
\bigl|\mathcal{M}_\mathrm{1st+2nd}(qg & \rightarrow qg)\bigr|^2 \\
=\frac{4}{9}g^4\times
\Biggl\{
	\frac{s}{M_\mathrm{s}^4}
	& \Biggl[\frac{s^2\bigl(-u\bigr)}
			{\bigl(s-M_\mathrm{s}^2\bigr)^2+\bigl(M_\mathrm{s}\Gamma^{J=1/2}_{q^*,\mathrm{1st}}\bigr)^2}
		+\frac{\bigl(-u\bigr)^3}
			{\bigl(s-M_\mathrm{s}^2\bigr)^2+\bigl(M_\mathrm{s}\Gamma^{J=3/2}_{q^*,\mathrm{1st}}\bigr)^2}
	\Biggr] \\
	\hspace{5mm}
	+\frac{s^3}{16M_\mathrm{s}^8}
	& \Biggl[\frac{1}{9}\frac{s^2\bigl(-u\bigr)}
			{\bigl(s-2M_\mathrm{s}^2\bigr)^2+\bigl(\sqrt{2}M_\mathrm{s}\Gamma^{J=1/2}_{q^*,\mathrm{2nd}}\bigr)^2}
		+\frac{9}{25}\frac{\bigl(-u\bigr)^3}
			{\bigl(s-2M_\mathrm{s}^2\bigr)^2+\bigl(\sqrt{2}M_\mathrm{s}\Gamma^{J=3/2}_{q^*,\mathrm{2nd}}\bigr)^2}
	\Biggr] \\
	\hspace{5mm}
	+\frac{s}{4M_\mathrm{s}^8}
	& \Biggl[\frac{1}{9}\frac{s^2\bigl(-u\bigr)\bigl(3t+s\bigr)^2}
			{\bigl(s-2M_\mathrm{s}^2\bigr)^2+\bigl(\sqrt{2}M_\mathrm{s}\Gamma^{J=3/2}_{q^*,\mathrm{2nd}}\bigr)^2}
		+\frac{1}{25}\frac{\bigl(-u\bigr)^3\bigl(5t+s\bigr)^2}
			{\bigl(s-2M_\mathrm{s}^2\bigr)^2+\bigl(\sqrt{2}M_\mathrm{s}\Gamma^{J=5/2}_{q^*,\mathrm{2nd}}\bigr)^2}
	\Biggr] \\
	\hspace{5mm}
	+\frac{s^2}{2M_\mathrm{s}^6}
	& \Biggl[\frac{1}{3}\frac{s^2\bigl(-u\bigr)}
			{\bigl(s-M_\mathrm{s}^2\bigr)^2+\bigl(M_\mathrm{s}\Gamma^{J=1/2}_{q^*,\mathrm{1st}}\bigr)^2}
		\frac{\bigl(s-M_\mathrm{s}^2\bigr)\bigl(s-2M_\mathrm{s}^2\bigr)
			+\sqrt{2}M_\mathrm{s}^2\,\Gamma^{J=1/2}_{q^*,\mathrm{1st}}\,\Gamma^{J=1/2}_{q^*,\mathrm{2nd}}}
			{\bigl(s-2M_\mathrm{s}^2\bigr)^2+\bigl(\sqrt{2}M_\mathrm{s}\Gamma^{J=1/2}_{q^*,\mathrm{2nd}}\bigr)^2} \\
		& \hspace{-2mm}
		+\frac{3}{5}\frac{\bigl(-u\bigr)^3}
			{\bigl(s-M_\mathrm{s}^2\bigr)^2+\bigl(M_\mathrm{s}\Gamma^{J=3/2}_{q^*,\mathrm{1st}}\bigr)^2}
		\frac{\bigl(s-M_\mathrm{s}^2\bigr)\bigl(s-2M_\mathrm{s}^2\bigr)
			+\sqrt{2}M_\mathrm{s}^2\,\Gamma^{J=3/2}_{q^*,\mathrm{1st}}\,\Gamma^{J=3/2}_{q^*,\mathrm{2nd}}}
			{\bigl(s-2M_\mathrm{s}^2\bigr)^2+\bigl(\sqrt{2}M_\mathrm{s}\Gamma^{J=3/2}_{q^*,\mathrm{2nd}}\bigr)^2}
	\Biggr] \\
	\hspace{5mm}
	+\frac{s}{M_\mathrm{s}^6}
	& \Biggl[\frac{1}{3}\frac{s^2\bigl(-u\bigr)\bigl(3t+s\bigr)}
			{\bigl(s-M_\mathrm{s}^2\bigr)^2+\bigl(M_\mathrm{s}\Gamma^{J=1/2}_{q^*,\mathrm{1st}}\bigr)^2}
		\frac{\bigl(s-M_\mathrm{s}^2\bigr)\bigl(s-2M_\mathrm{s}^2\bigr)
			+\sqrt{2}M_\mathrm{s}^2\,\Gamma^{J=1/2}_{q^*,\mathrm{1st}}\,\Gamma^{J=3/2}_{q^*,\mathrm{2nd}}}
			{\bigl(s-2M_\mathrm{s}^2\bigr)^2+\bigl(\sqrt{2}M_\mathrm{s}\Gamma^{J=3/2}_{q^*,\mathrm{2nd}}\bigr)^2} \\
		& \hspace{-2mm}
		+\frac{1}{5}\frac{\bigl(-u\bigr)^3\bigl(5t+s\bigr)}
			{\bigl(s-M_\mathrm{s}^2\bigr)^2+\bigl(M_s\Gamma^{J=3/2}_{q^*,\mathrm{1st}}\bigr)^2}
		\frac{\bigl(s-M_\mathrm{s}^2\bigr)\bigl(s-2M_\mathrm{s}^2\bigr)
			+\sqrt{2}M_\mathrm{s}^2\,\Gamma^{J=3/2}_{q^*,\mathrm{1st}}\,\Gamma^{J=5/2}_{q^*,\mathrm{2nd}}}
			{\bigl(s-2M_\mathrm{s}^2\bigr)^2+\bigl(\sqrt{2}M_\mathrm{s}\Gamma^{J=5/2}_{q^*,\mathrm{2nd}}\bigr)^2}
	\Biggr] \\
	\hspace{5mm}
	+\frac{s^2}{4M_\mathrm{s}^8}
	& \Biggl[\frac{1}{9}\frac{s^2\bigl(-u\bigr)\bigl(3t+s\bigr)}
			{\bigl(s-2M_\mathrm{s}^2\bigr)^2+\bigl(\sqrt{2}M_\mathrm{s}\Gamma^{J=1/2}_{q^*,\mathrm{2nd}}\bigr)^2}
		\frac{\bigl(s-2M_\mathrm{s}^2\bigr)^2
			+2M_s^2\,\Gamma^{J=1/2}_{q^*,\mathrm{2nd}}\,\Gamma^{J=3/2}_{q^*,\mathrm{2nd}}}
			{\bigl(s-2M_\mathrm{s}^2\bigr)^2+\bigl(\sqrt{2}M_\mathrm{s}\Gamma^{J=3/2}_{q^*,\mathrm{2nd}}\bigr)^2} \\
		& \hspace{-2mm}
		+\frac{3}{25}\frac{\bigl(-u\bigr)^3\bigl(5t+s\bigr)}
			{\bigl(s-2M_\mathrm{s}^2\bigr)^2+\bigl(\sqrt{2}M_\mathrm{s}\Gamma^{J=3/2}_{q^*,\mathrm{2nd}}\bigr)^2}
		\frac{\bigl(s-2M_\mathrm{s}^2\bigr)^2
			+2M_\mathrm{s}^2\,\Gamma^{J=3/2}_{q^*,\mathrm{2nd}}\,\Gamma^{J=5/2}_{q^*,\mathrm{2nd}}}
			{\bigl(s-2M_\mathrm{s}^2\bigr)^2+\bigl(\sqrt{2}M_\mathrm{s}\Gamma^{J=5/2}_{q^*,\mathrm{2nd}}\bigr)^2}
	\Biggr]
\Biggr\} \,.
	\end{split}
\end{equation}


\end{document}